\documentclass[twocolumn]{aa} 

\usepackage{amsmath}
\usepackage{graphicx}

\usepackage{txfonts}
\usepackage{epstopdf}
\usepackage{siunitx}
\newcommand\Gaia{\textit{Gaia}}

\newcommand{\unit}[1]{\ensuremath{\, \mathrm{#1}}}
\begin{document}

   \title{Gaia Data Release 1}

   \subtitle{On-orbit performance of the Gaia CCDs at L2}
   \author{C.~Crowley\inst{\ref{inst:esac_cmc}} \and R.~Kohley\inst{\ref{inst:esac}} \and N.C.~Hambly\inst{\ref{inst:edinb}} \and M.~Davidson\inst{\ref{inst:edinb}} \and A.~Abreu\inst{\ref{inst:esac_asier}} \and F.~van~Leeuwen\inst{\ref{inst:cambridge}}
\and C.~Fabricius\inst{\ref{inst:ieec}} \and G.~Seabroke\inst{\ref{inst:mssl}} 
\and J.H.J.~de~Bruijne\inst{\ref{inst:estec}}
\and A.~Short\inst{\ref{inst:estec}}
\and L.~Lindegren\inst{\ref{inst:lund}}
\and A.G.A.~Brown\inst{\ref{inst:leiden}}
\and G.~Sarri\inst{\ref{inst:estec}}
\and P.~Gare\inst{\ref{inst:estec}}
\and T.~Prusti\inst{\ref{inst:estec}}
\and T.~Prod'homme\inst{\ref{inst:estec}}
\and A.~Mora\inst{\ref{inst:auroraesac}} 
\and J.~Mart\'in-Fleitas\inst{\ref{inst:auroraesac}}
\and F.~Raison\inst{\ref{inst:esac_fred},\ref{inst:max_planck}}
\and U.~Lammers\inst{\ref{inst:esac}} \and W.~O'Mullane\inst{\ref{inst:esac}}
\and F.~Jansen\inst{\ref{inst:estec}}
}

\institute{
HE Space Operations BV for ESA/ESAC, Camino Bajo del Castillo s/n, 28691 Villanueva de la Ca{\~n}ada, Spain \\ \email{cian.crowley@esa.int}
\label{inst:esac_cmc}
\and
ESA, European Space Astronomy Centre, Camino Bajo del Castillo s/n, 28691 Villanueva de la Ca{\~n}ada, Spain
\label{inst:esac}
\and
Institute for Astronomy, School of Physics and Astronomy, University of Edinburgh, 
Royal Observatory, Blackford Hill, Edinburgh, EH9~3HJ, UK,
\label{inst:edinb}
\and
Deimos-Space S.L.U.\ for ESA/ESAC, Camino Bajo del Castillo s/n, 28691 Villanueva de la Ca{\~n}ada, Spain 
\label{inst:esac_asier}
\and
Institute of Astronomy, University of Cambridge, Madingley Road, Cambridge CB3 0HA, UK,
\label{inst:cambridge}
\and
Dept. d'Astronomia i Meteorologia, Institut de Ci\`encies del Cosmos,
Universitat de Barcelona (IEEC-UB), Mart\'i Franqu\`es 1, E-08028 Barcelona, 
Spain 
\label{inst:ieec}
\and
Mullard Space Science Laboratory, University College London, Holmbury St Mary, Dorking, Surrey RH5 6NT, UK 
\label{inst:mssl}
\and
ESA, European Space Research and Technology Centre, Keplerlaan 1, 2200 AG, Noordwijk, The Netherlands
\label{inst:estec}
\and
Lund Observatory, Department of Astronomy and Theoretical Physics, Lund University, Box 43, 22100, Lund, Sweden
\label{inst:lund}
\and
Sterrewacht Leiden, Leiden University, P.O.\ Box 9513, 2300 RA, Leiden, The Netherlands
\label{inst:leiden}
\and
Aurora Technology for ESA/ESAC, Camino Bajo del Castillo s/n, 28691 Villanueva de la Ca{\~n}ada, Spain
\label{inst:auroraesac}
\and
Praesepe for ESA/ESAC, Camino Bajo del Castillo s/n, 28691 Villanueva de la Ca{\~n}ada, Spain
\label{inst:esac_fred}
\and
Max Planck Institute for Extraterrestrial Physics, OPINAS, 
Giessenbachstrasse, 85741 Garching, Germany
\label{inst:max_planck}
}
 
   \date{ }

\abstract{
The  European Space Agency's Gaia satellite  was launched into  orbit around L2 in December 2013  with a payload containing 106 large-format scientific CCDs.
The primary goal of the mission is to repeatedly obtain high-precision astrometric and photometric measurements of one thousand million stars over the course of five years.
The scientific value of the down-linked data, and the  operation of the onboard autonomous detection chain,  relies on the high performance of the detectors. As Gaia slowly rotates and scans the sky, the CCDs are  continuously operated in a mode where the line clock rate and the satellite rotation spin-rate are in synchronisation.
Nominal mission  operations began in July 2014 and the first data release is being prepared for release at the end of Summer 2016.

In this paper we present an overview of the focal plane, the detector system, and strategies for on-orbit performance monitoring of the system. This is followed by a presentation of the  performance results based on analysis of data acquired during a two-year window beginning at payload switch-on. Results for  parameters such as readout noise and electronic offset behaviour  are presented and  we pay particular attention to the effects of the L2 radiation environment on the devices.
The radiation-induced degradation in the charge transfer efficiency (CTE) in the (parallel) scan direction is clearly diagnosed; however, an extrapolation shows that charge transfer inefficiency (CTI) effects at end of mission will be approximately an order of magnitude less than predicted pre-flight. 
It is shown that the CTI in the serial register (horizontal direction) is still dominated by the traps inherent to the manufacturing process and that the radiation-induced degradation so far is only a few per cent. We also present results on the tracking of ionising radiation damage and hot pixel evolution.
Finally, we summarise some of the  detector effects discovered on-orbit which are still being investigated.

}


   \maketitle

\section{Introduction}

The European Space Agency's Gaia spacecraft was launched into orbit in December 2013.  Nominal operations were entered in July 2014 and the satellite will continue to operate at the Earth/Moon--Sun Lagrangian point for the duration of the nominal mission lifetime of 5 years. 
Rotating slowly, {\Gaia} scans the sky so that its two optical telescopes  repeatedly observe more than one thousand million stars, as well as  many galaxies, quasars, and solar system objects. The resulting data set will be iteratively reduced to solve for the  position, parallax, and proper motion of every observed star. The final data release will also include photometry, colours, low-resolution spectra, and astrophysical parameters of every star, along with radial velocities and medium-resolution spectra of the brighter objects. For details on the mission, see Gaia collaboration (2016a) in this
volume\footnote{For regular updates on the science performances and the mission in general see http://www.cosmos.esa.int/web/gaia.}. Also published in this volume is the first intermediate data release (Gaia collaboration 2016b) which is based on data obtained during the first year of operations.

The quality of the released data will depend on the performance of the more than one hundred large-format, custom-built  CCD detectors operating on the focal plane. Indeed, for the science goals to be met, the devices must operate correctly and within specifications until the  end of the mission. Of particular concern pre-flight was the effect of the radiation environment at L2 on the CCD performances, in particular with respect to charge transfer efficiency (CTE) degradation induced by non-ionising damage from  protons (mostly).
In  this paper we present an overview of the characteristics and operational parameters of the devices, discuss the methods used to monitor on-orbit performance and obtain calibrations, and present results on the  measured performance parameters and radiation diagnostics over the course of the first two years of the mission.

\section{Hardware description}
\subsection{ Gaia focal plane\label{sect:fpa}}

\begin{figure*}[htb]
        \begin{center}
        \includegraphics[width=\textwidth]{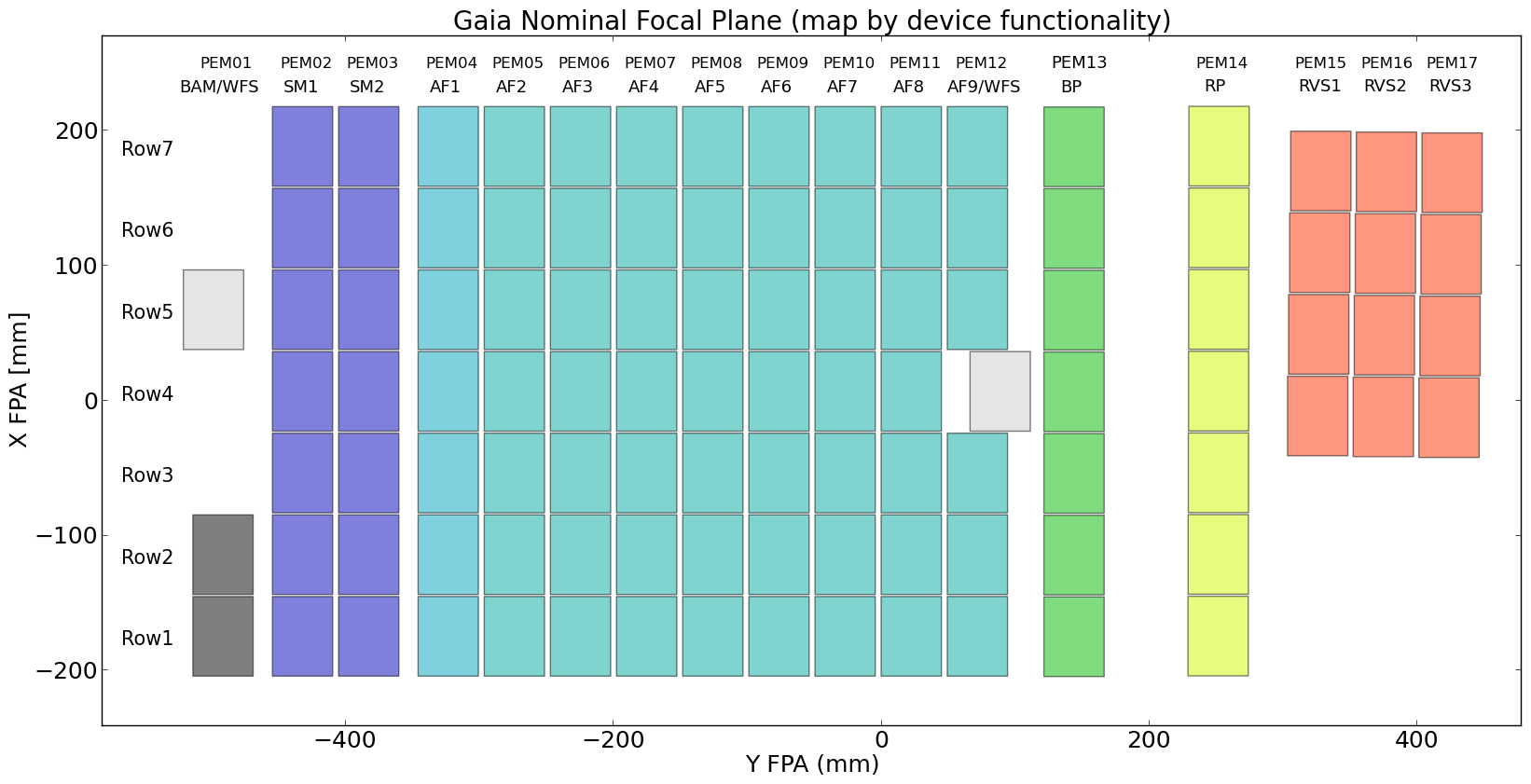}
        \caption{\label{fig:fpa} The  Gaia focal plane (FPA) consists of 106 large-format CCDs arranged in seven rows, where all CCDs on a particular row are operated by a dedicated onboard computer, all seven in synchronisation. We thus designate each of these rows as CCD Row~$1 - 7$, as denoted in the schematic above. In the horizontal direction, the CCD array is split into 17 different  CCD strips (from CCD strip `BAM/WFS' to `RVS3'). The CCDs are colour-coded according to the functional group that they fall into. The light grey colour is used to denote Wavefront Sensor devices (WFS), the darker grey for the Basic Angle Monitor (BAM) instrument, dark blue for the devices used for onboard detection (Sky Mappers CCDs: SM), light blue for the AF (Astrometric Field) CCDs, green for the BP (Blue Photometer), yellow for the RP (Red Photometer), and red for the RVS (Radial Velocity Spectrometer) CCDs. As can be observed, all CCD rows are not identical, but include different functionalities. For example, some rows contain RVS CCDs to collect object spectra during star transits, and some do not. All rows,  however, do contain all of the common elements required for the autonomous detection, confirmation, astrometric measurement, and photometric measurement chain: SM, AF1, AF2-9 (commonly grouped as `AF'), BP, and  RP CCDs. The total height of the FPA from CCD row 1 to 7 corresponds to the scan width of about 0.7$^{\circ}$ on the sky.
}
        \end{center}
\end{figure*}

The Gaia focal plane array (FPA) consists of 106 large area CCDs mounted on a silicon-carbide support structure. The  detectors are continuously operating in time delay and integration (TDI) mode where the line\footnote{In Gaia parlance, a row of pixels is often called a line to distinguish it from a `row' of CCDs in the FPA.} transfer rate and the satellite rotation spin-rate are in synchronisation.   See Figure~\ref{fig:fpa} for a schematic of the FPA where the CCDs are colour-coded based on their general function; we note that the transit direction is from left to right (see the caption for further details).  

Taking the case of a  transit\footnote{The direction of transit is called the along-scan (AL) direction in Gaia parlance; this corresponds to what is typically called the parallel (or vertical) direction in CCD terminology. The serial (or horizontal) direction is usually referred to as the across-scan (AC) direction.} of a typical star across one CCD row of the FPA, the process starts with autonomous onboard detection in one of the SM CCDs. The telescopes share a common FPA, so baffling is used to ensure that light from telescope 1 is detected in SM1 and light from telescope 2 is detected by the SM2 CCD. Of course, the SM CCDs must be operated in full-frame mode in order to permit the unbiased onboard detection  \citep[for details on the onboard process see][]{2015AA...576A..74D}. After a star transits along an SM CCD and is read out and if the star is correctly detected, a window (whose dimensions and on-chip binning scheme depend on the measured brightness) is assigned in order to measure the predicted transit along the AF1 CCD. Since most sources are faint, the large majority of observations are binned~$\times12$~pixels in the AC direction for AF observations. After window readout, onboard algorithms determine whether the initial detection has been confirmed or not, and if the result is positive, then further windows are assigned to the rest of the AF CCDs on that row in order to carry out lower-noise (see Table~\ref{tab:noiseTab}), high-precision astrometric measurements. In addition, extended windows  are also assigned to the BP and RP CCDs which act as photometric devices as they are located under  blue- (BP) and red- (RP) optimised prisms,  respectively.

If the source is bright enough, and the row contains RVS CCDs (i.e. CCD rows~$4 - 7$), then windows are also  assigned in order to track the predicted transit over these three   CCDs. These CCD transits  acquire medium-resolution spectral measurements centred  on a wavelength region around the CaII triplet in the near infrared. There are also two sets of two CCDs which are set aside for the BAM device and WFS measuring instrument. For further general information on the FPA, see \cite{2012SPIE.8442E..1PK}. For a report on the  on-orbit performance during payload commissioning of the BAM and WFS instruments, see \cite{2014SPIE.9143E..0XM}.

\subsection{ CCDs}
\label{sect:ccds}

\begin{figure}[htb]
        \begin{center}
                \includegraphics[width=\columnwidth]{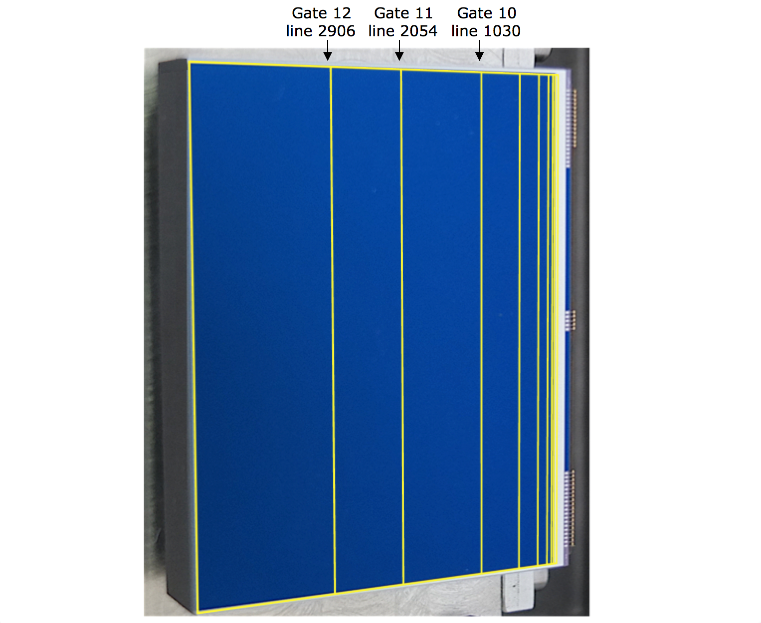}
        \caption{\label{fig:ccd}  Gaia CCD with the position of the special TDI gates overlayed in yellow. The charge injection structure is to the left and the serial register to the right. Only the three TDI gates closest to the charge injection structure are marked. Towards the serial register the spacing between gates is successively reduced by a factor of 2 and becomes too closely spaced for illustration. The image orientation follows the Gaia convention with the scan direction in the image area from left to right. The physical dimensions of each device are approximately $4.7\times 6.0$~\rm{cm}~(AL~$\times$~AC).
}
        \end{center}
\end{figure}

The Gaia CCDs (designated  CCD91-72) were custom  designed and  manufactured by e2v technologies. They are back-illuminated full-frame devices specifically made for the Gaia TDI mode of operation. Each device contains  a series of design peculiarities and additional features, amongst which are rectangular pixels, a charge injection structure for radiation-induced CTI mitigation and diagnosis, a lateral anti-blooming drain for blooming control, a supplementary buried channel (or `notch'), a two-phase serial register, and special TDI blocking gates to reset the TDI integration time, resulting in one of the most complex devices ever made by e2v. Some features are described below; a picture of the CCD is shown in Figure~\ref{fig:ccd}.

The image area consists of 4500 $\times$ 1966 rectangular four-phase pixels of physical dimension 10 $\times$~\SI{30}{\micro\metre}, which translates into~$\sim59 \times 177$~milliarcseconds on the sky. The pixel size is a compromise between the required resolution  in the along-scan (AL) direction and full well capacity.
The TDI line clock rate is just under~\SI{1}{\milli\second} (\SI{0.9828}{\milli\second}),  which results in an effective exposure time of~\SI{4.42}{\second}  for a CCD transit.

To ease operational complexity and owing to the implementation of a single output node, there is a single two-phase serial register at the end of the image area which contains 1966 pixels plus 15 lead-in pixels for  prescan acquisition. To reduce jitter between charge transfer in the image area and the movement of the star images along the CCD, and unlike in traditional imaging mode, the parallel clocking has to be interleaved with the serial readout of a single  line of pixels. The parallel binning required for some operations modes can therefore not be done directly into the serial register, but into a separately clocked  line of pixels with increased full well capacity between the image area and the serial register which temporarily holds the binned charge until the next serial register readout starts. Serial binning is done directly onto the output node.

The devices were manufactured in three variants, each one with identical chip architecture, but   optimised to improve their quantum efficiency for specific wavelength regimes:

\begin{itemize}

\item Most  devices (78 CCDs) mounted on the single integrated FPA  were designed for the purpose of  broad-band astrometric measurements and are called the AF variant. They are built on standard silicon thinned to approximately \SI{16}{\micro\metre}  and have a broad-band single-layer AR-coating centred at 650nm. They are used for the science instrument FPA functions SM and AF and the auxiliary WFS instrument function.

\item The blue photometer (BP) science instrument integrated in the FPA required CCDs most sensitive in the blue, achieved through an enhanced surface passivation process and the AR-coating centred at 360nm. They are called the BP variant and only the seven blue photometer CCDs on the FPA are of this variant.

\item The red photometer (RP) and also the radial velocity spectrometer (RVS) science instruments integrated in the FPA required CCDs more sensitive towards redder wavelengths, called the RP variant. They were therefore fabricated on high-resistivity, deep-depleted silicon thinned to approximately \SI{40}{\micro\metre}  and have an AR-coating centred at 750nm. In addition to  the seven RP and 12 RVS CCDs, this variant is also used for the two BAM CCDs on the FPA. 

\end{itemize}

The fundamental difference of standard silicon used for the AF and BP variant versus high-resistivity silicon for the RP variant together with increased device thickness leads to different device performances like charge transfer inefficiency (CTI) between these two populations of variants, as can be seen in later sections.

\subsubsection{Pixel architecture}
\label{sect:pixel}

The pixel structure is based on a standard buried channel implant, standard oxide, and four overlapping polysilicon gates. An additional doping of smaller lateral extent is added to the buried channel forming a narrower potential maximum or notch for signal transport. This structure is called the supplementary buried channel (SBC) and confines small signal charge packets to a smaller transport volume and therefore improves CTI for faint objects. For details on the Gaia SBC design and testing  of pre-flight efficacy, see \citet{2013MNRAS.430.3155S}. In addition to the SBC running along all pixels of each column a lateral anti-blooming drain (ABD) is implanted adjacent to the traditional channel stops that isolate the individual columns. The function of the lateral anti-blooming structure is to avoid vertical blooming by draining the excess charge of saturated objects during TDI integration before it can spill into neighbouring pixels. The potential on the drain is set such that excellent charge transfer efficiency is obtained up to saturation and full well becomes only limited by non-linearity. For details on the pixel architecture see \citet{2010SPIE.7742E..14S} and \citet{2005SPIE.5902...31S}.

\subsubsection{TDI blocking gates}
\label{sect:gates}

Single polysilicon gates on 12 of the 4500 pixel rows have separate electrical contacts to act as TDI blocking gates. While in nominal TDI operation the corresponding polysilicon gates of each pixel row in the four-phase design are driven synchronously, these special gates can be held at a barrier potential, which blocks the transport of charge within the column at that point. Accumulated excess charge is drained to the ABD until the gate is released and again synchronously clocked in TDI mode. The effect is that the star image moves on, but the accumulated charge in the CCD is no longer transported in synchronisation but stopped, and therefore the effective TDI integration time restarts at the position of the activated gate. Activation of one of the 12 gates, based on the prediction of a bright star transit along the CCD, can keep the accumulated signal for bright objects below the saturation level. The effective TDI integration times can be adjusted from the nominal 4.42s down to approximately 2ms (factor of over 2000) leading to an enormous dynamic range for Gaia observations.

The gate for 2ms integration time is closest to the serial register and leaves only two pixel rows active for exposure. Permanently activating this gate basically stops all charge accumulating in the image area from reaching the serial register and is used in a special onboard calibration procedure to assess the bias (or electronic offset) along the CCD line readout.

\subsubsection{Charge injection structure}
\label{sect:ciStructure}

Charge injection is a method used to fill pixel potentials with electronically generated charge. At the top of the image area adjacent to pixel row 4500 (farthest from the serial register)  is an injection structure that consists of an injection drain and a single injection gate. The injection level is set by an adjustable potential difference between the pixel potential on pixel row 4500 and the injection gate potential and injection itself is controlled in a fill-and-spill method by applying a clock pulse to the injection drain. This design and operation is inherently radiation-hard with very uniform and stable injection level over the injected lines within the same column and over time, but unfortunately non-uniform across the columns. The injection level can be tuned from zero up to saturation level, but the spatial non-uniformity leads to injection into only a few columns for the lowest injection levels. The ideal injection level, amount of injected lines, and the frequency of injection for nominal operation has been assessed during extensive on-ground radiation tests.  The current baseline is about 20ke$^{-}$ of injection level, four consecutive injection lines injected every few seconds (the period is fixed but different for the different instruments). Like the TDI gate activation, charge injection is fundamental to Gaia CCD operation on-orbit and used for calibration, radiation damage monitoring, and mitigation purposes and is referred to throughout the remainder of the paper.

\subsubsection{Detector readout}
\label{sect:readoutChain}

While the architecture and interfaces of all Gaia CCDs are identical, the operating modes of each FPA function and instruments are quite different. For example, the BAM CCDs integrate a static interference pattern during exposure and need to operate in stare mode in addition to the TDI readout. In addition,  the serial readout frequencies, on-chip binning, and system gain settings are widely different between the individual readout modes. Instead of designing dedicated CCD controller systems for each CCD function in the FPA, a single common set of readout electronics was custom built by Crisa under contract by EADS Astrium (now Airbus Defence \& Space) and known as Proximity Electronic Modules (PEM). The different Gaia CCD operating modes are all programmable via software. The increased design complexity is counter-balanced by production advantages, exchangeability, and increased on-orbit operational flexibility, the last most important for on-ground and on-orbit analysis and mitigation of detected operational anomalies. Each PEM is paired with a single flight-CCD in the FPA. A special command package is sent to the PEM from the  onboard computer in order to initialise the operational mode. This is followed by a command packages that is sent every TDI line period to set readout parameters such as readout window location, binning, charge injection, and TDI gate activation. 

An overview of the different operating modes and selected characteristics is given in Table~\ref{tab:romodes}. The characteristics of readout frequency, readout mode, binning, and gain are a trade-off to achieve the required signal-to-noise ratios and dynamic range for the number of windows (or full-frame) and number of pixels per window required for each FPA function given the constraint of the fixed TDI line time of about 1ms for all detectors on the FPA.

\begin{table}[!h]
  \caption{Selected operating mode characteristics for the science instruments nominal CCD modes.}
  \label{tab:romodes}
\begin{center}
  \begin{tabular}{  c | c | c | c}
    \hline
    CCD mode  & Readout & Readout & Binning  \\
                       & frequency & mode &  (Ser.~$\times$~Par.) \\
    \hline
    SM & 833kHz & Full Frame & 2~$\times$~2  \\
    AF1 & 400kHz & Window & 2~$\times$~1  \\ 
   AF & 103kHz & Window & 1~$\times$~1, 12~$\times$~1  \\ 
    BP & 164kHz & Window & 1~$\times$~1, 12~$\times$~1  \\
    RP & 164kHz & Window & 1~$\times$~1, 12~$\times$~1  \\  
    RVS-HR\footnotemark & 172kHz & Window & 1~$\times$~1, 10~$\times$~1 \\ 
     &   &   &   \\
     \hline
   CCD mode  & System & Charge & TDI gate  \\
                       & gain & injection & activation  \\
    \hline
    SM & low & no & permanent \\
    AF1 & low & yes & yes \\ 
    AF & low & yes & yes \\ 
    BP & low & yes & yes \\
    RP & low & yes & yes \\  
    RVS-HR & high & no & no \\ 
    \hline
  \end{tabular}

\end{center}
\end{table}
\footnotetext{It was originally planned to also use a low-resolution RVS mode (RVS-LR) for acquisition of fainter spectra. However, its use  was discarded at the end of the commissioning phase owing to some small thermal instabilities introduced into the payload as a result of the constant sky-dependent reconfiguring of the detectors.}

The PEM analogue-digital chain is based on digital correlated double sampling (DCDS) with multiple sampling and averaging in the digital space to reduce readout noise. The excellent noise performance with respect to the operating mode of Table~\ref{tab:romodes} can be seen in Table~\ref{tab:noiseTab}.
The performance of the detector systems on-orbit has been excellent as well with not a single failure so far.

\section{On-orbit detector performance}
\label{sect:performances}

\subsection{Detector calibration and performance monitoring strategies}
\label{sect:strategy}

Owing to the scan/survey nature of the mission and the TDI operation of the CCDs, the Gaia detector system is not equipped with a shutter or an onboard lamp, thus rendering some conventional detector calibration  procedures impossible (i.e. standard dark image acquisitions, or using a lamp for gain change monitoring or for non-linearity calibrations, flatfields, etc.). In any case, many of these effects, which would traditionally require stand-alone calibrations, can instead be calibrated out using the actual science data within  the iterative global solution \citep[][]{2012A&A...538A..78L}. At this relatively early stage into the mission (and into the iterative data processing cycles)  the extraction of some of these parameters from the science data is rather complicated. For example, any possible detector gain or quantum efficiency  changes  would be  convolved with the optical throughput changes induced by the mirror-contamination issue (Fabricius et al., this volume). Therefore, in this paper we do not address gain, quantum efficiency, non-linearity, or saturation parameters and instead we focus on those performance parameters which are directly measurable (for an overview of the detector performances during the commission phase see \citealt{2014SPIE.9154E..06K}).

Despite the self-calibrating philosophy of Gaia, some detector parameters  do require dedicated measurements, and these are carried out using one or more of three strategies:

\begin{enumerate}

\item Virtual objects (VOs). These are special windows which are acquired by inserting artificial detections into the  algorithms of the onboard computers, thus propagating  a virtual stellar object along the focal plane and acquiring the resulting windowed image from each CCD. The size and binning of the window can be configured to  one of the window sizes possible for nominal observations.  During normal operations, special patterns are continuously run, providing a steady stream of information from every science CCD. During the detection process, most of these objects are assigned lower priority than real sky windows,  so during times of  high star density on the focal plane, fewer VOs will be acquired. These windows provide useful information that can be utilised to monitor levels of stray light, transient feature rates, defect column evolution,  and the behaviour of the detector electronic offset levels in the image area. Importantly, these patterns sample each CCD column and also sample the charge injection and release trail, thus providing a way of monitoring the evolution of the radiation-induced degradation of the CTE in the AL direction, this is discussed in some detail in Section~\ref{sec:al_cti}.

\item Special calibration activities. For some calibration acquisitions, the nominal operation of the whole row of CCDs needs to be stopped and  special sequences to be run in order to acquire the required information from the devices. Currently four detector-related special calibration activities are periodically executed on board. These are  related to  monitoring and characterising of non-uniformities in the electronic offset level during readout  (Section~\ref{sec:non_uninformity}), AL and AC CTI effects (Sections~\ref{sec:al_cti} and \ref{sec:ac_cti}), and the accumulated effects of ionising radiation (Section~\ref{sec:ionising_radiation}). Indeed, some of these activities involve the strategic placement of VOs in order to correctly sample each CCD.

\item Prescan acquisitions.   Prescan pixels are acquired from  every CCD during readout. However,  the packetised science data are stripped of these data, so in order to monitor changes in the zero-point of the electronic offset, $1024$ prescan pixels  from each CCD are periodically stored in a high downlink-priority packet providing  periodic `bursts' of information. For an analysis of these data, see Sections~\ref{sec:prescan} and \ref{sec:noise}. 

\end{enumerate}

After two years at L2, generally speaking, all devices are functioning and performing well. The onboard detection and confirmation chain, using the SM and AF1 CCDs is working as expected at the detector level and the on-orbit measured performance parameters correspond well with the on-ground measured values. In the remainder of this section we examine in  greater detail some of these performance parameters over the course of the mission so far. To place these results into context with respect to the temperature dependency of these performance parameters, we first provide an overview of the temperature evolution of the FPA.

 \subsection{FPA temperature stability}
\label{sec:temperature}

\begin{figure}
        \begin{center}
        \includegraphics[width=\columnwidth]{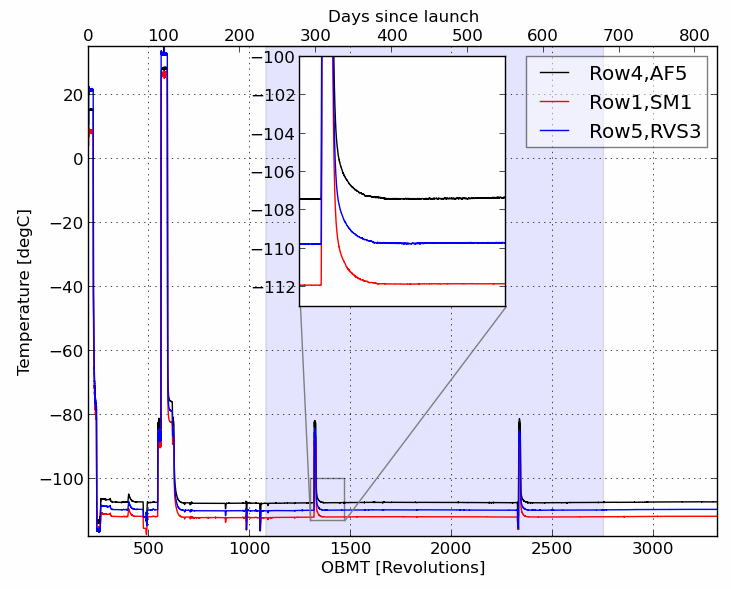}
        \caption{\label{fig:temperature} Readings for the three temperature sensors closest to the FPA. The legend shows the names of the CCDs which are closest to each of the three sensors. We note the large temperature increases due to mirror decontamination events and also the offset between the sensor readings highlighting the temperature variation over the focal plane (see text). Shaded in blue  is the time period when data was acquired, which is used for the first data release. The beginning of this time period corresponds to the end of the commissioning phase where there were many temperature changes made during the configuration of the payload (testing the effect of heaters on the payload stability etc.).
}
        \end{center}
\end{figure}

Amongst the many different temperature sensors on board the satellite, there are three placed close to the detector array. They are distributed across the FPA with one placed near the SM1 CCD in row 1, one placed near the AF5 CCD in row 4, and the third  near the RVS3 CCD in row 5 (see Figure~\ref{fig:fpa} for a map of the FPA detector array).  Shown in Figure~\ref{fig:temperature} is a plot of the readings obtained from these three sensors.
Three very large temperature increases which take place after the initial cool-down phase are clearly visible. These correspond to controlled heating events which were carried out in order to decontaminate the mirrors (Gaia collaboration 2016a, this volume). It can be observed that the first decontamination was quite aggressive and resulted in room  temperatures being reached which are typically high enough to achieve an anneal of some radiation damage effects  \citep[e.g.][]{2009stis.rept....2G,2005ITNS...52.2672M}. It is not anticipated that the Gaia CCDs will be annealed in the future, but we note that this is possible if required. Any future nominal heating events will display a signature similar to the two most recent heating events that are visible on the plot.
The top axis on this plot displays the number of days since Gaia launch and   the bottom axis shows the equivalent values, but displayed in a unit of spacecraft revolutions that  we call Onboard Mission Timeline (OBMT). The orbital spin period of the satellite is approximately six hours, so there are four OBMT revolutions in a $24$-hour period. The convention chosen was to place the zero-point in this scale $50$~days ($200$~revolutions) before launch. We use this unit throughout the remainder of this paper when referring to the mission timeline.

It is apparent that when thermal equilibrium is reached after each heating event the FPA temperature  is extremely stable. Large temperature changes can dramatically affect  detector characteristics (such as CTI and gain), so long-term stability is an important aspect of the performance. There is an obvious gradient of a few degrees over the focal plane, but this remains stable, and the operating temperatures of all devices are close to the target value of  $163.15$~K.
 \subsection{Prescan stability}
\label{sec:prescan}

The zero-point electronic offset of the CCDs is expected to vary throughout  the mission, especially when the payload is subjected to temperature changes. The monitoring of these detector offsets is carried out using data down-linked in special prescan packets, as discussed in Section~\ref{sect:strategy}. Every line read out from  each  of the CCDs will sample one unbinned and one binned prescan pixel (the exact binning scheme depends on the CCD operating mode), with the exception of the SM CCDs where all  $14$ accessible prescan pixels are read out (hardware binned~$\times~2$).

\begin{figure}
        \begin{center}
                \includegraphics[width=\columnwidth]{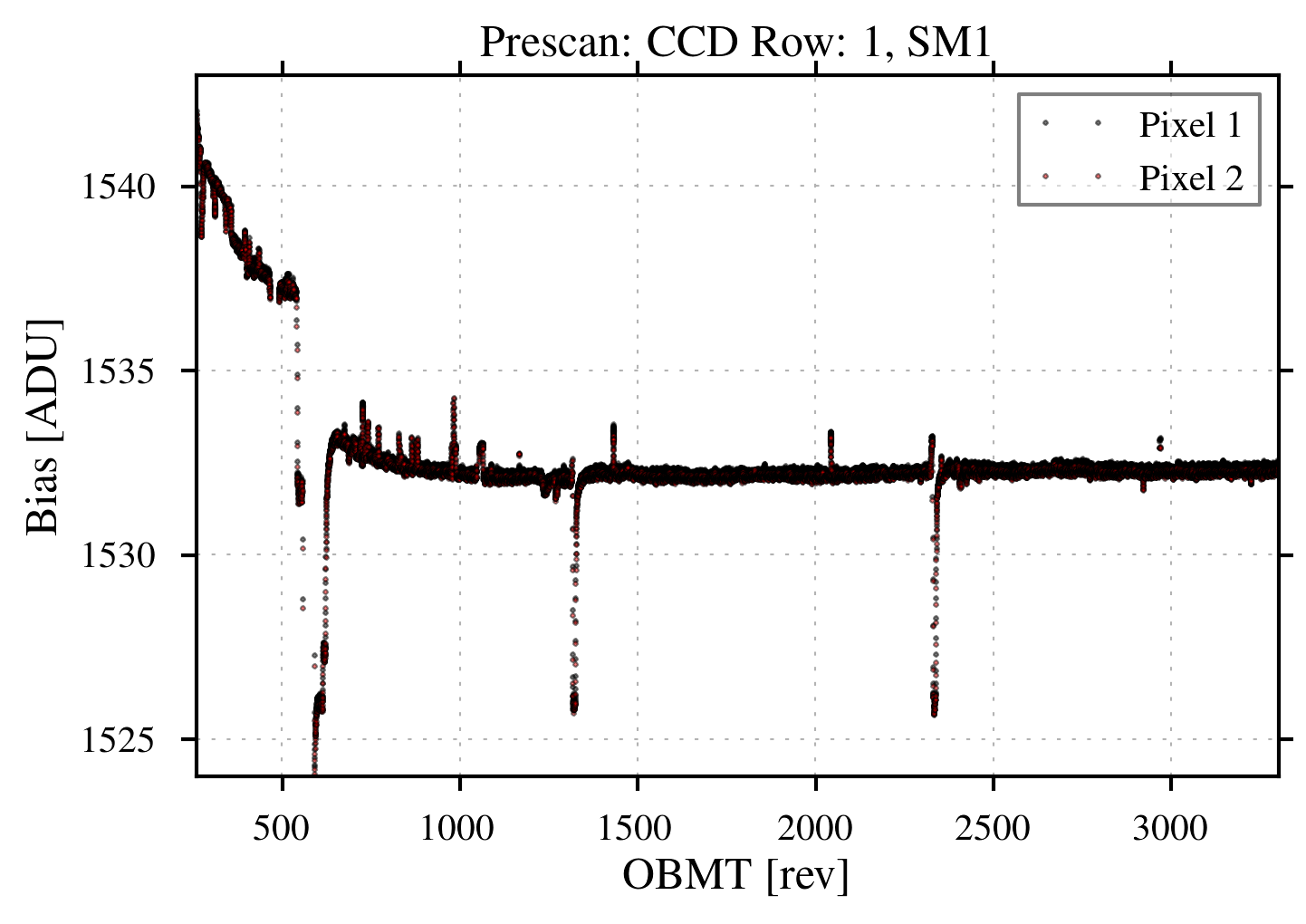}
                \caption{\label{fig:prescan} 
 Long-term trend of the first two prescan pixels from the SM1 CCD in row 1. It is important to note the correlation of variability with payload heating events (see Figure~\ref{fig:temperature}) and the relatively stable behaviour during periods of thermal stability. There are also some short-term increases in the bias levels which are not correlated with temperature changes. The origin of these spikes is currently being investigated.
}
        \end{center}
\end{figure}

An examination of the evolution of the prescan values\footnote{The electronic offset varies within a line from one pixel to another (see Section~\ref{sec:non_uninformity} for more details).} shows that the behaviour is generally stable over time, with the exception of variations around times of the controlled payload heating events, for example, see  Figure~\ref{fig:prescan}. Some devices show some long-term trends 
and the cause of the variations is still being investigated; however, the zero-point offset can be successfully tracked for all devices and used for calibration purposes.

 \subsection{CCD noise performance}
\label{sec:noise}

Using the same prescan data as were discussed in the previous section,  the short-term variability can be used to estimate and monitor the readout noise for each device. Owing to the different functionalities of each CCD mode, each has different trade-off characteristics in order  to be compliant with the TDI line time. Therefore, each also has different noise specifications. For example, low noise is required for AF and RVS-HR mode devices in order to maximise astrometric precision and low-noise spectral measurement, whereas the prime purpose of the SM CCDs is detection, so these devices need to be read in full-frame mode requiring fast pixel sampling, and the noise requirements are thus necessarily relaxed (see Table~\ref{tab:romodes} for further details).

\begin{figure}
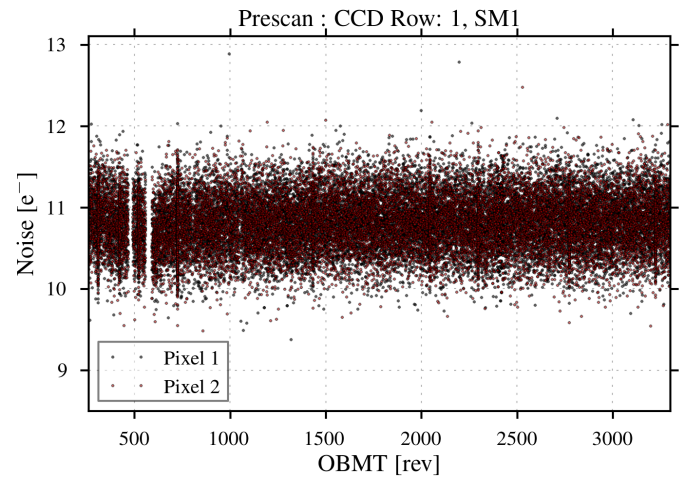

        \begin{center}
\includegraphics[width=\columnwidth]{{{figures/PrescanNoiseCCDRow1SM1}.png}}
                \caption{\label{fig:noise} 
 Long-term noise profile of  two prescan pixels acquired from the SM1 CCD in row 1. 
}
        \end{center}
\end{figure}

In general, the noise performance is excellent and stable on-orbit  for all devices. Shown in Figure~\ref{fig:noise} is the long-term noise behaviour for an example  CCD. The  level is extremely stable.  Owing to the difficulty in directly measuring the transfer gain on-orbit, the on-ground measured gain values are used in order to  express the results in units of electrons. The high observed stability of the  noise therefore also suggests that, so far, there are no perceived gain changes due to radiation damage.

\begin{figure} 
        \begin{center}
        \includegraphics[width=\columnwidth]{{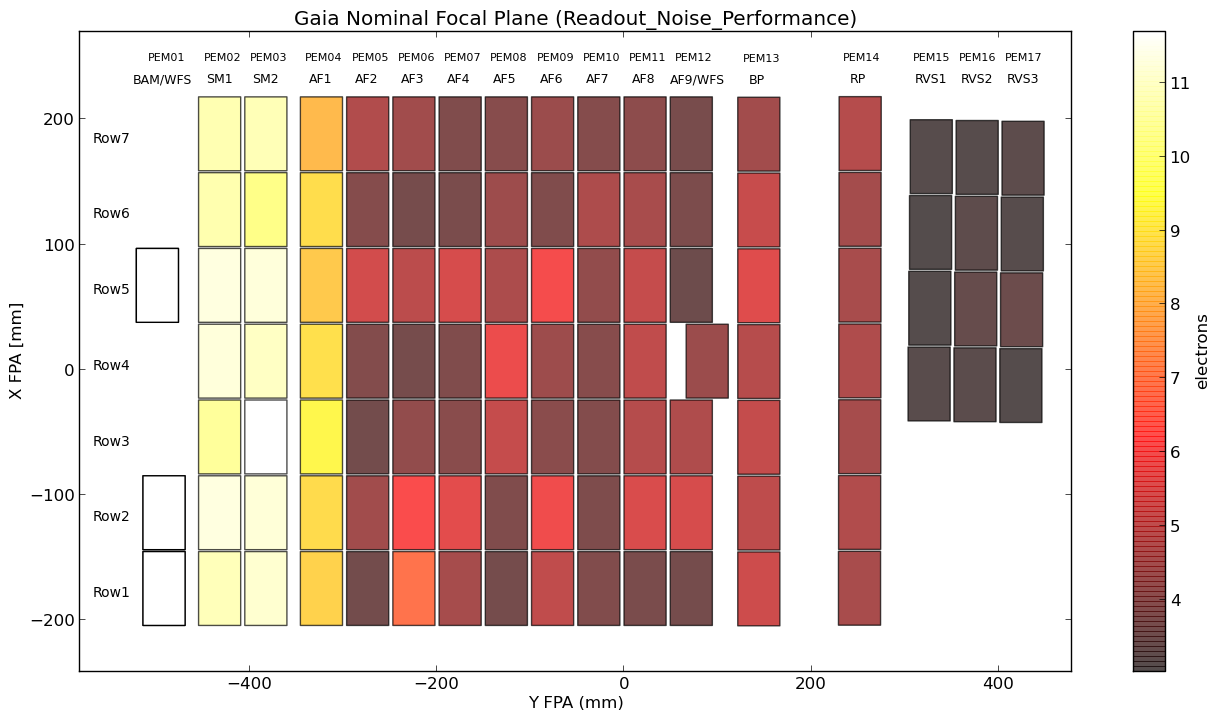}}
        \caption{\label{fig:noise_dist} 
Distribution of readout noise measurements for all science CCDs using data acquired between OBMT revolutions $1500$ and $2000$. The highest noise levels are in the SM CCD strips, where the full-frame readout required for autonomous detections means that the noise requirements need to be relaxed. The detection confirmation step is carried out using the AF1 mode CCDs and $162$~pixels need to be  read out per line, so the noise is a little lower.  Overall, the noise performances can be seen to be excellent (see text and Table~\ref{tab:noiseTab} for further details).
}
        \end{center}
\end{figure}

For some devices there is some small long-term variation in the measured noise values, sometimes present for one prescan pixel and not the other, but this is  a minor effect and only present for some devices. Shown in Figure~\ref{fig:noise_dist} is the distribution of readout noise measurements across the FPA for all science CCDs using data acquired between OBMT revolutions $1500$ and $2000$ when the temperatures were stable.  It should be noted that most devices perform much better than the specifications, with only one device (the AF3 CCD in row 1) falling slightly outside.  

Presented in  Table~\ref{tab:noiseTab} are the mean and standard deviations of the measured on-orbit noise parameters per CCD operating mode compared to the pre-flight specifications. It can be observed that, overall, the on-orbit noise performance is excellent.

\begin{table}[!h]
  
\caption{Measured readout noise results for the Gaia science CCDs measured between OBMT revolutions $1500$ and $2000$. The mean noise and $\sigma_{\text{noise}}$ parameters refer to the mean and standard deviations of the values of the readout noise measured for all devices operating in that mode.}
\label{tab:noiseTab}
\begin{center}
  \begin{tabular}{  c | c | c | c}
    \hline
    CCD mode  & Specification   & Mean noise & $\sigma_{\text{noise}}$   \\ 
       & \unit{e^{-}} & \unit{e^{-}} & \unit{e^{-}} \\ \hline
    SM & $<$13 & 10.99 & 0.39 \\ 
    AF1 & $<$10 & 8.77 & 0.36 \\ 
    AF & $<$6.5 & 4.48 & 0.79 \\ 
    BP & $<$6.5 & 5.08 & 0.35 \\
    RP & $<$6.5 & 4.66 & 0.11 \\  
    RVS-HR & $<$6 & 3.18 & 0.13 \\ 
    \hline
  \end{tabular}
\end{center}

\end{table}

 \subsection{Bias non-uniformity}
\label{sec:non_uninformity}

The use of prescan measurements (as detailed in Section~\ref{sec:prescan}) to remove 
the electronic offset in image area pixels   presupposes that this
level does not change during the readout of the serial register. 
However, this is not the case. In fact, there
is a set of complex `bias non-uniformities' present in the offsets that need
to be modelled  in order to correctly remove the bias level from a given pixel.

\begin{figure}
\centerline{\includegraphics[width=\columnwidth,clip]{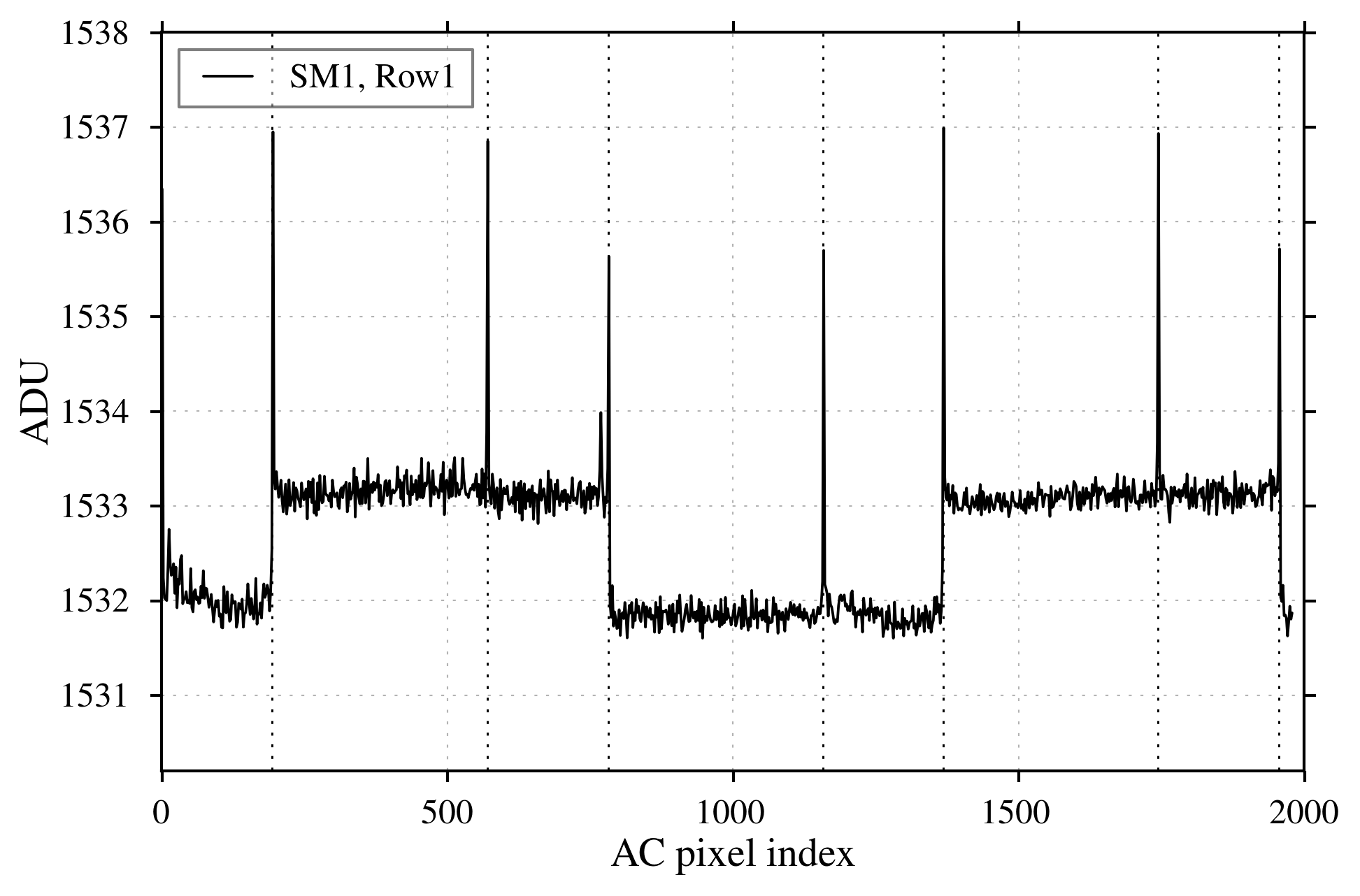}}
\caption[]{ AC bias profile of the SM1 CCD in row 1 (with the TDI gate closest to the serial register raised), the dashed vertical lines denote the pixels that are affected by the glitch anomaly described in the text.  \label{fig:glitches}}
\end{figure}

The description of the origin, calibration, and mitigation of all these bias
non-uniformities is beyond the scope of this paper and will be reported
elsewhere in the future. In summary, however, we note  that the operation of the serial
register is interrupted and paused four times between each cycle of the interleaved parallel clocks in order
to avoid a large offset perturbation resulting from the  clock swings
of the four-phase CCDs (without the pauses the perturbation is of the order of $70$~ADU). Furthermore, for windowed readout modes (i.e.~all
modes except SM which employs full-frame readout) the unwanted pixels between
windows are fast-flushed to enable more time to be spent digitising the window
pixels with a consequent reduction in read noise. However, perturbations to image area pixel values (relative to the 
prescan  pixel  values) are observed, with amplitudes in the range~1 to~100~analogue-to-digital units (ADU) depending on
device and mode. These perturbations are repeatable, and therefore amenable to
calibration, but they have rather complex dependencies on the  timing of the readout.
To illustrate this point, shown in  Figure~\ref{fig:glitches} is the AC bias profile of the SM1 CCD in row 4. In SM mode no pixels are flushed;  however, a number of sharp jumps in the bias level, as well as baseline offset changes, can easily be observed. The sharp jumps in the offset level correspond to pixels that are read out immediately after resumption of the serial readout after one of the pauses has been applied.  This effect is known as the   `glitch' anomaly. 

\begin{figure}
\centerline{\includegraphics[width=\columnwidth,trim=0 25 0 25,clip]{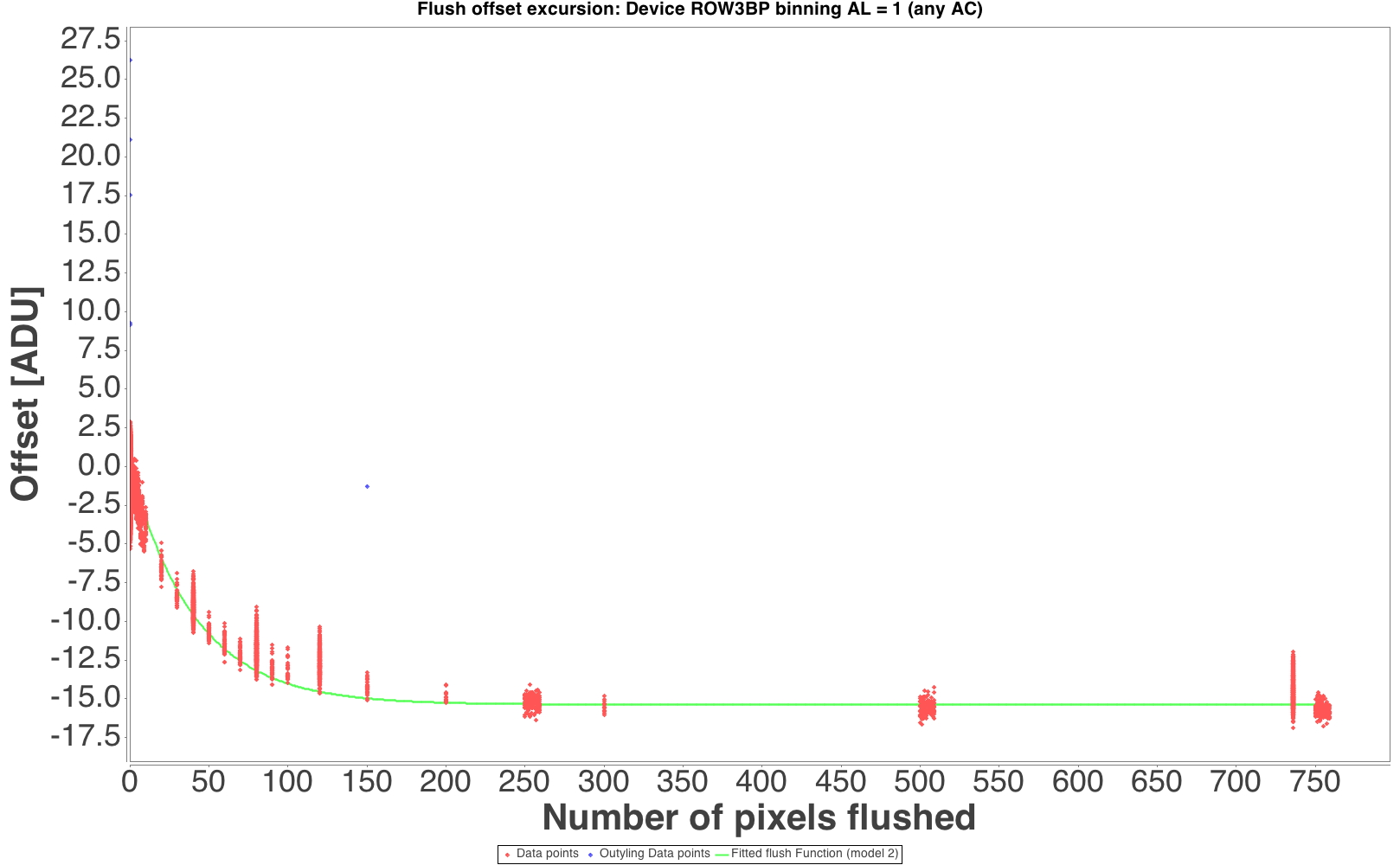}}
\caption[]{Electronic offset dependency on the number of pixels fast-flushed immediately before readout. These data are from the BP CCD in row 3.  \label{fig:flush}}
\end{figure}

An example of the dependency of the offset level change on the number of pixels fast-flushed immediately before readout is shown in
Figure~\ref{fig:flush}.

There are at least four components to this non-uniformity, so in order to disentangle and calibrate the various components of the non-uniformity, a special calibration activity is periodically run on board. This activity consists of activating various carefully designed VO patterns across the FPA that allow the calibration of each component of the non-uniformity. To date, the activity has been run nine times and each run has generally been successful (see Section~\ref{sec:investigation} for an outline of some outstanding issues). 

 \section{Effects of the L2 radiation environment}
\label{sec:radiation}

The Sun-Earth Lagrange point L2 offers many  advantages for the operation of astronomical observatories, such as high thermal and gravitational stability. Owing to the large distance of L2 from Earth ($\sim1.5$~million km), the effect of geomagnetically trapped charged particles is not a concern. However, this also implies that L2 lacks the shielding of the Earth’s magnetosphere and is thus vulnerable to impacts from ionised particles from other sources. Indeed, the L2 radiation environment is dominated by particles (mostly protons) from effectively isotropic Galactic cosmic rays (GCRs) and  the more directional solar eruptive events. The GCR component is expected to be rather steady throughout the mission, with the number of impacts expected to vary smoothly and to be anti-correlated with the solar cycle (see Section~\ref{sec:ppe_detection}). In contrast, the impacts of particles from Earth-directed solar events are sporadic with more events expected to be observed around the time of solar maximum. 

Generally speaking, the most damaging effects of radiation on CCDs can be divided into two categories:  1) total ionising dose (TID) and 2) performance degradation from displacement damage (also referred to as non-ionising energy loss, or NIEL). 
Total ionising dose effects   occur when a charge accumulates in parts of some electronic devices, which can result in threshold shifts, charge leakage, etc. A special calibration procedure is periodically run on board to monitor the accumulation of charge in the oxide layers of the CCD detectors; this is  described further in Section~\ref{sec:ionising_radiation}.

Non-ionising energy
loss effects are probably those most likely to interfere with the scientific objectives of the Gaia mission. The effects are  cumulative and occur as a result of bulk damage to the silicon crystalline structure of the CCDs, usually caused by protons. Spacecraft shielding is predicted to protect against the lower energy particles ($\lesssim 1$~MeV), whilst the probability of displacement damage from high-energy particles decreases towards high energies.  Therefore, it is expected that the majority of the damage will occur from impacting particles with energies approximately in the range   $1-100~\unit{MeV}$. 
Radiation-induced defects can damage pixels and subsequently result in new hot or defective pixels and an increase in dark current rates (see Section~\ref{sec:defects}). However, of greater  concern for Gaia, is electron trapping and subsequent delayed release of charge by  new energy levels (or electron `traps') during the transfer of photoelectrons from pixel-to-pixel (see Sections~\ref{sec:al_cti} and \ref{sec:ac_cti}).

Whilst not physically damaging to the  detectors, transient events also affect the Gaia data. These effects are the result of  collisions between incident charged particles and electrons in the device silicon which result in the creation of electron-hole pairs producing  characteristic `track' features in the images.  These features can pollute the downlinked windowed-images and need to be accounted for in the on-ground processing. Indeed, an analysis of these features can provide useful information on the changing radiation environment (see Section~\ref{sec:transients}). Also, owing to the detection process  carried out on board, these transient events need to be filtered out in real-time. An analysis of the statistics of these transient feature rejections is presented in the following section.

\subsection{Onboard PPE Counters}
\label{sec:ppe_detection}

The impact of both GCRs and solar particles on the CCDs  causes transient features, or prompt-particle events (PPE), to be read out from the detectors.
In order to discriminate non-star-like objects from stars, the object detection algorithms running on board scan the images read out from each SM CCD in search of local flux maxima where low and high spatial frequency filters are applied. This autonomous onboard detection-rejection process is configured via a set of onboard parameters that were fine-tuned during the commissioning phase in order to achieve an adequate balance between detection performance requirements and rejection. Amongst these `rejected'  objects will be many image ghosts, extended objects, etc.  A special algorithm also rejects features which are deemed too sharp to have been convolved with the PSF from the optics, i.e. typical PPE features.  Statistics on this  process are collected and telemetered  to the ground in the form of auxiliary science telemetry. An analysis of PPE counter data provides insight into the instantaneous radiation environment experienced by Gaia at L2.  We note that the onboard detection is working well at the detector level; however, there are  many cases of  spurious detections, especially around bright stars (Fabricius et al., this volume).

\begin{figure}[!htbp]
        \begin{center}
        \includegraphics[width=0.5\textwidth,height=0.28\textheight]{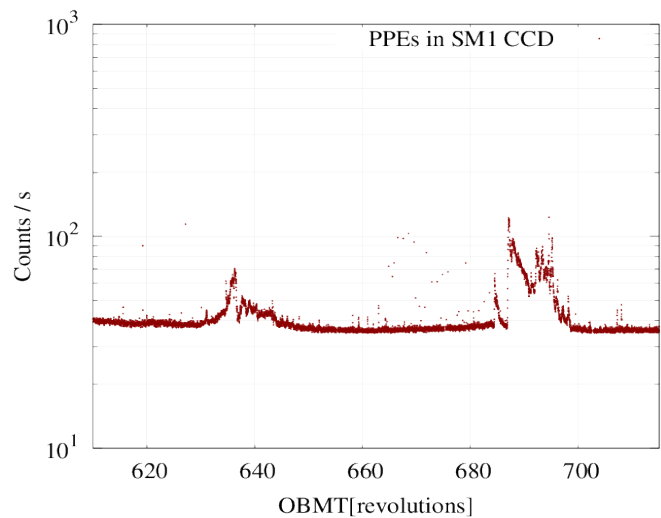}
        \caption{ PPE count rates detected on board in the SM1 CCD.\label{fig:ppes}}
        \end{center}
\end{figure}

Figure~\ref{fig:ppes} shows the typical PPE count rates measured by one SkyMapper CCD for a particular period during the early Gaia mission. We can clearly distinguish  two main features:

\begin{enumerate}
        \item `Bursts' of increased radiation, see e.g. the peaks at OBMT$\sim$640 and $\sim$690 revolutions;
        \item A `background' (pedestal) count rate (at ${\sim}40$ counts$~\rm{s}^{-1}$ in the figure)  present at all times, on top of which increased radiation activity shows up.
\end{enumerate}

Following \citet{sullivan1971geometric}, the PPE rate measured at each CCD (SM or AF) produced by an incident isotropic flux of cosmics is given by 

\begin{equation*}
        \begin{aligned}[c]
        \rm R_{CCD} = \frac{F_{CCD} \cdot A }{2} counts \cdot s^{-1},
        \end{aligned}
 \end{equation*}

where \rm{A} is the effective
detection area of the device in cm$^2$ (A$\simeq$17.1 cm$^2$ for SM CCDs) and $\rm{F}$ is the particle flux measured in particles cm$^{-2}$~\rm{s}$^{-1}$.  Assuming a typical  particle background at L2 of 5 protons cm$^{-2}$~\rm{s}$^{-1}$ \citep[][]{catalano2014characterization}, the expected PPE rate measured  by Gaia at L2 is expected to be $\sim42.8$ counts~\rm{s}$^{-1}$, which is in good agreement ($<$10\% difference) with in-flight measured rates. 

\begin{figure}[!h]
        \begin{center}
        \includegraphics[width=0.5\textwidth,height=0.27\textheight]{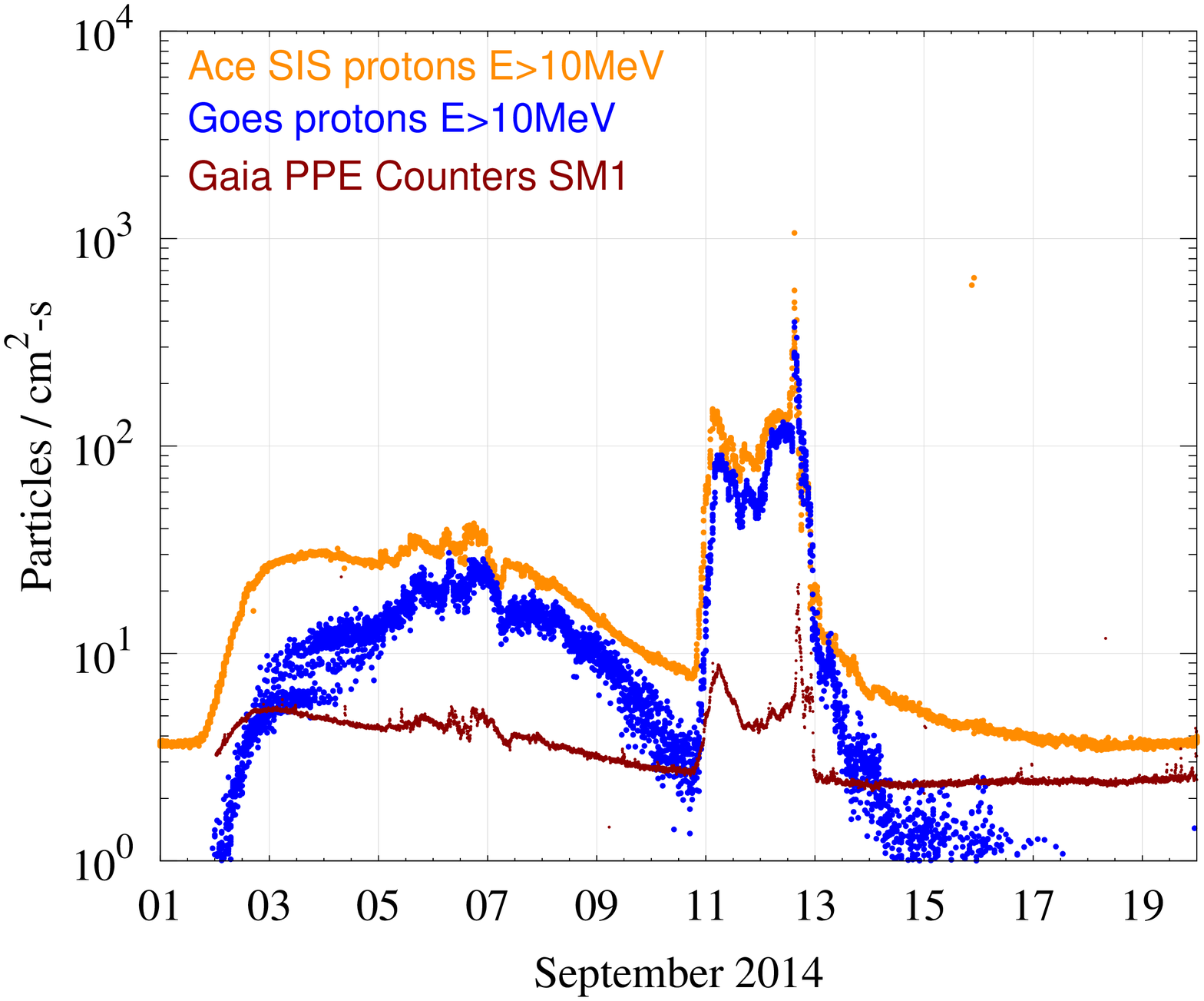}
        \includegraphics[width=0.5\textwidth,height=0.27\textheight]{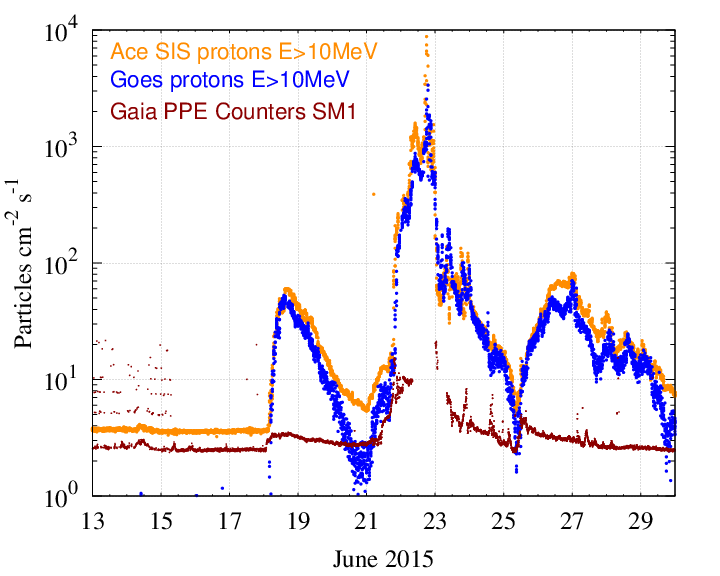}
        \caption{ {\bf{Top: }}  Data showing a comparison between  NASA's ACE (orange) SIS Instrument, Goes15 EPAM (blue), and Gaia (dark red) cosmic ray fluxes comparison for the September 2014  X1 solar flare event.  {\bf{Bottom: }} Similar to the top panel, but for the M6 June 2015 event.\label{fig:comparison}}
        \end{center}
\end{figure}

In Figure~\ref{fig:comparison} we compare the Gaia counter data acquired around the times of two clear bursts to data from other spacecraft radiation monitoring instruments  over the same timescales in order to search for correlations. Indeed, an examination of the comparison plots shows that the correlation of the bursts with solar events is clear.

\subsubsection{Long-term PPE counter behaviour}

\begin{figure}[!h]
        \begin{center}
        \includegraphics[width=0.5\textwidth,height=0.28\textheight]{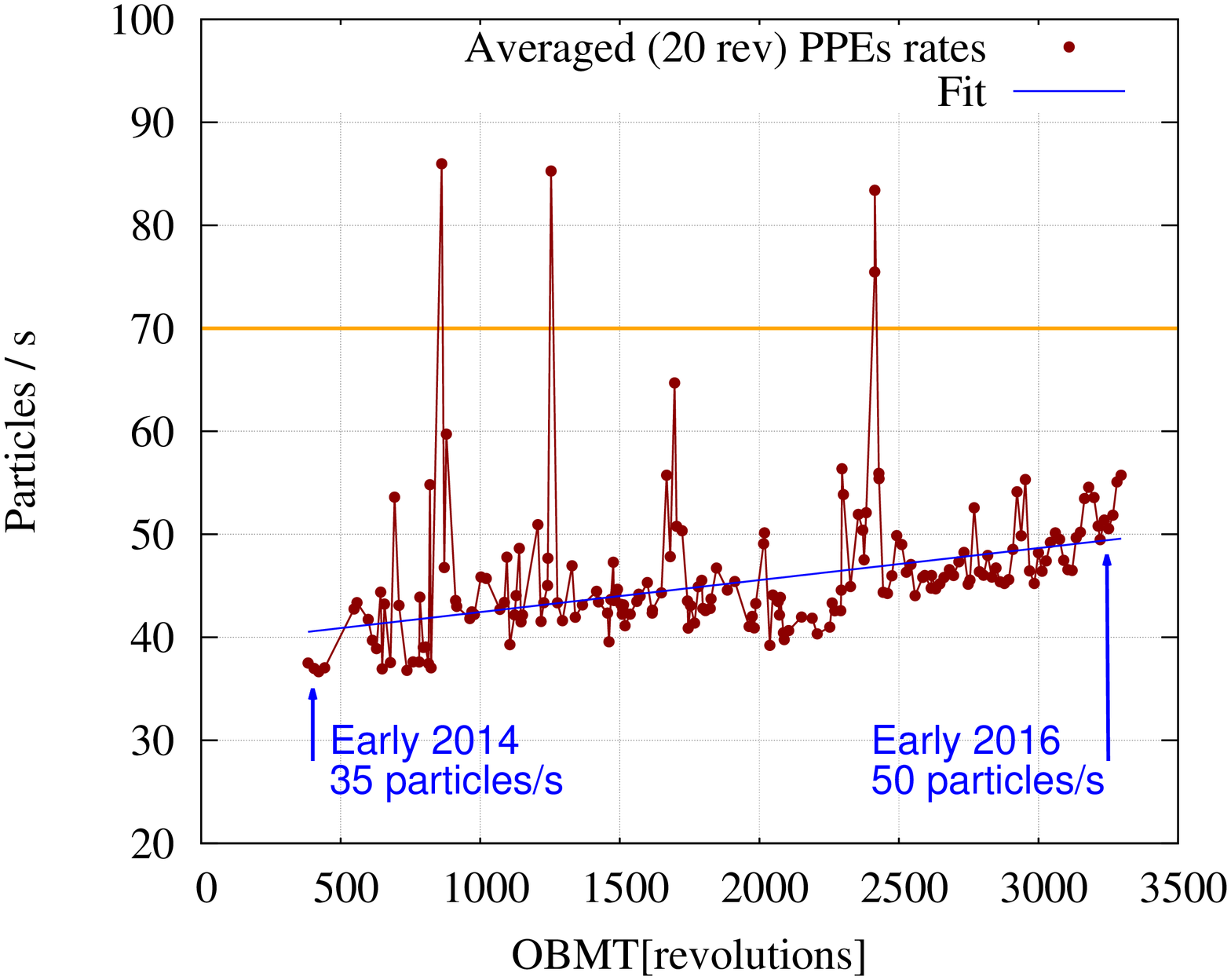}
        \includegraphics[width=0.5\textwidth,height=0.28\textheight]{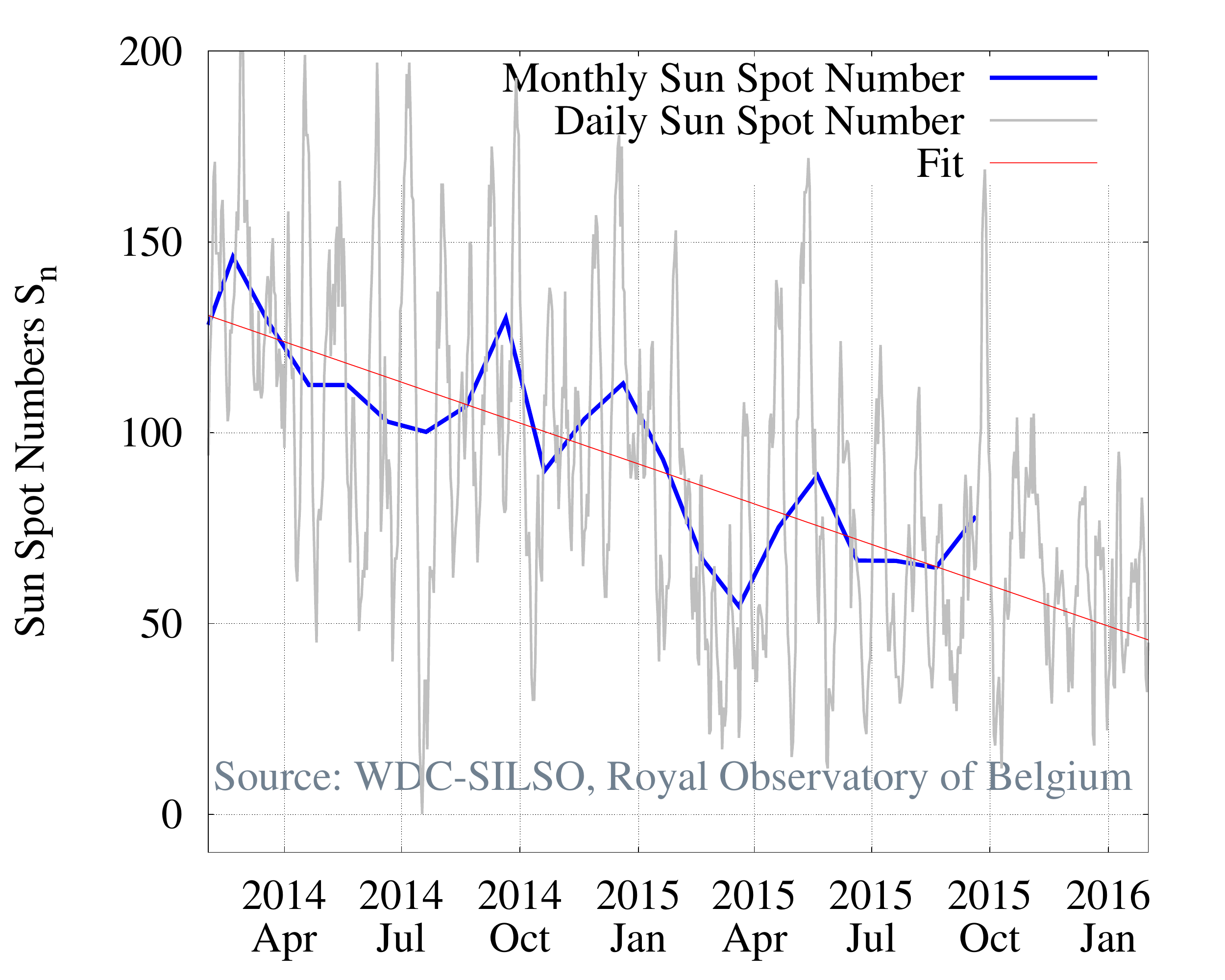}
        \caption{Long-term trend of Gaia PPE counter statistics averaged for all the SM CCDs (using a four-satellite revolution bin size) showing a slow but steady increase in the pedestal from early 2014 to early 2016 (top) and historical 2014-2016 solar activity measured by daily and monthly Sun spot numbers (bottom).\label{fig:long-term}}
        \end{center}
\end{figure}

An analysis of  the long-term trend of the  PPE counter statistics shows that the pedestal value has evolved from a value of  ${\sim}38$ counts$~\rm{s}^{-1}$ in early 2014 to a value of ${\sim}50$ counts$~\rm{s}^{-1}$ in early 2016. This steady increase in the PPE rates measured by Gaia at L2, contrasts with the steady  decrease in solar activity since  launch via Sun spot numbers (see Figure~\ref{fig:long-term}). It has long been known \citep{fisk1969solar} that solar activity modulates the GCR fluxes with a decreased GCR rate observed towards solar maximum and an increased GCR rate towards solar minimum. In the Gaia case, the spacecraft arrived at L2 coinciding with the start of a decreasing solar activity cycle after a maximum in December 2013.

The observed long-term trend in the PPE counter data appears to generally follow this anti-correlation with solar activity and would therefore suggest that the pedestal in PPE counters corresponds to GCRs which are continuously `bathing' L2.  On top of this pedestal the solar radiation particle interactions show up as bursts.

 \subsection{Transients}
\label{sec:transients}

Whilst it is important that transient features are filtered out of the onboard detection chain, it has always been known that some fraction of down-linked stellar images (and VOs) will be polluted by these features. Indeed, an analysis of the count rates  of VOs polluted by transient features over the course of the mission shows a pattern very similar to that observed in the PPE count data, albeit with poorer statistics. In order to improve on the statistics, an analysis of the 2D science windows for the AF1--AF9 CCDs was carried out over a time interval of $\sim17$~days. This interval covers the time of the X1 solar proton event of September 2014 described in the previous section. This  was carried out by using standard techniques to separate sharp, high-contrast cosmic hits from the lower contrast features (stars, etc.) that are affected by the  point spread function (PSF).

It was found that the occurrence of transient features in the windows science data over this period is relatively low ($\sim 0.16\%$) despite the presence of the solar flare event in the data. During the two days after the event approximately $1\%$ of windows were affected, although we note that these estimates depend strongly on the detection thresholds used in the analysis. Indeed, the proton event is very clearly observed in the data, both in terms of the count rates over time and also in terms of the amount of energy deposited into the detectors. Beginning approximately five hours after the event was noted on the sun,  a period of increased activity lasting about  two days was observed in the deduced transient rates. In addition, during this two-day period, the average maximum flux of the events gradually increased from $\sim~7500$~ADU to a maximum value of  $\sim 20000$~ADU, before eventually falling back to normal values ($\sim 6500$~ADU).

Interestingly, the effect  of the spin-dependent variation in the shielding against radiation can be observed in the transient rate data. The spatial distribution of the count rates across the focal plane shows a pattern that agrees very closely with the spatial distribution of the radiation damage across the FPA from  the same proton event; this will be discussed further in Section~\ref{sec:al_cti}. In addition, the number of transients observed during the solar proton event is a function of the rotation phase of the spacecraft. A similar effect is observed in the PPE count data and both of these observations can be explained by the fact that the effective  shielding that protects each CCD depends on the orientation of the spacecraft with respect to the direction of the incoming solar particles.

 \subsection{CTI in the scan direction (AL)}
\label{sec:al_cti}

Prior to launch it was recognised that untreated CTI effects in the AL (parallel) direction would have the potential to contribute significantly to the systematic errors in the final Gaia catalogue. Because pre-flight AL CTI are very low, most  scan-direction CTI effects  will be caused by radiation-induced defect sites in the CCD silicon lattice structure. These sites result in the  trapping and release of electrons during the TDI transfer of the integrating charge packet. For some Gaia-specific pre-flight studies on the effects of radiation on the science data see e.g. \citet[][]{2012MNRAS.419.2995P, 2012MNRAS.422.2786H,2005ITNS...52.2664H}. 

The most important hardware mitigation for AL CTI effects is the periodic injection of charge into the devices\footnote{The TDI gate 12 is permanently raised during nominal operation of the SM CCDs in order to permanently reduce the effective TDI integration time. This means that charge injection cannot be used for these devices since the injected charge would simply accumulate behind the raised gate potential and never propagate the full length of the CCD. For the nominal operation of the RVS CCDs,  no charge injection is used  since the charge in the leading edge of each spectrum will fill many empty traps and thus mitigate the effects (the  larger window length of the RVS spectra  would also mean that more RVS observations would be polluted by the injected charge). It is not used in WFS nor BAM CCDs either.}. This is carried out every $\sim 2$~seconds for AF1 and AF2-9 devices and every $\sim 5$ seconds for BP and RP devices. This scheme sees four contiguous lines of charge  injected into the CCD and clocked out through the $4500$~rows (see Section~\ref{sect:ciStructure}). This has a number of benefits:

\begin{itemize}

\item The periodic filling of traps with injected electrons, thus ensuring that there will be fewer active trapping sites available when packets of photoelectrons are traversing the detector. For those traps with characteristic release timescales on the order of seconds,  these traps will be kept filled over the period of the charge injection,  keeping them effectively permanently filled and thus rendered effectively inactive.
\item The resetting of the illumination history. It can be assumed that the four lines of charge will momentarily fill enough active traps to reset the illumination history for each CCD column (i.e. star transits prior to the charge injection do not need to be modelled).  In order to effectively model the CTI effects in the on-ground data treatment it is necessary to  include the effects of recent transits of other stars across the particular CCD columns in question. With charge injection implemented, it follows that  only the transits that have occurred since the last injection need to be accounted for. 

\item The regular presence of the injection features in the data stream means that the CTI effects on the injections can be used to monitor the evolution of the trapping and release effects on the detector over time. This manifests itself through two diagnostics. The first pixel response (FPR)  is measured by computing the number of electrons removed from the first injected line through trapping (in practice it is necessary to monitor the charge removed from all injected lines). The other diagnostic involves the measurement of the charge in the release trail behind the injected lines caused by the release of those electrons trapped by the FPR. See Figure~\ref{fig:testbench} for an example of trapping and release effects on a block of lines of injected charge. These data were acquired on ground on a proton-irradiated Gaia engineering model CCD (see the caption for further detail).

\begin{figure}
        \begin{center}
        \includegraphics[width=\columnwidth]{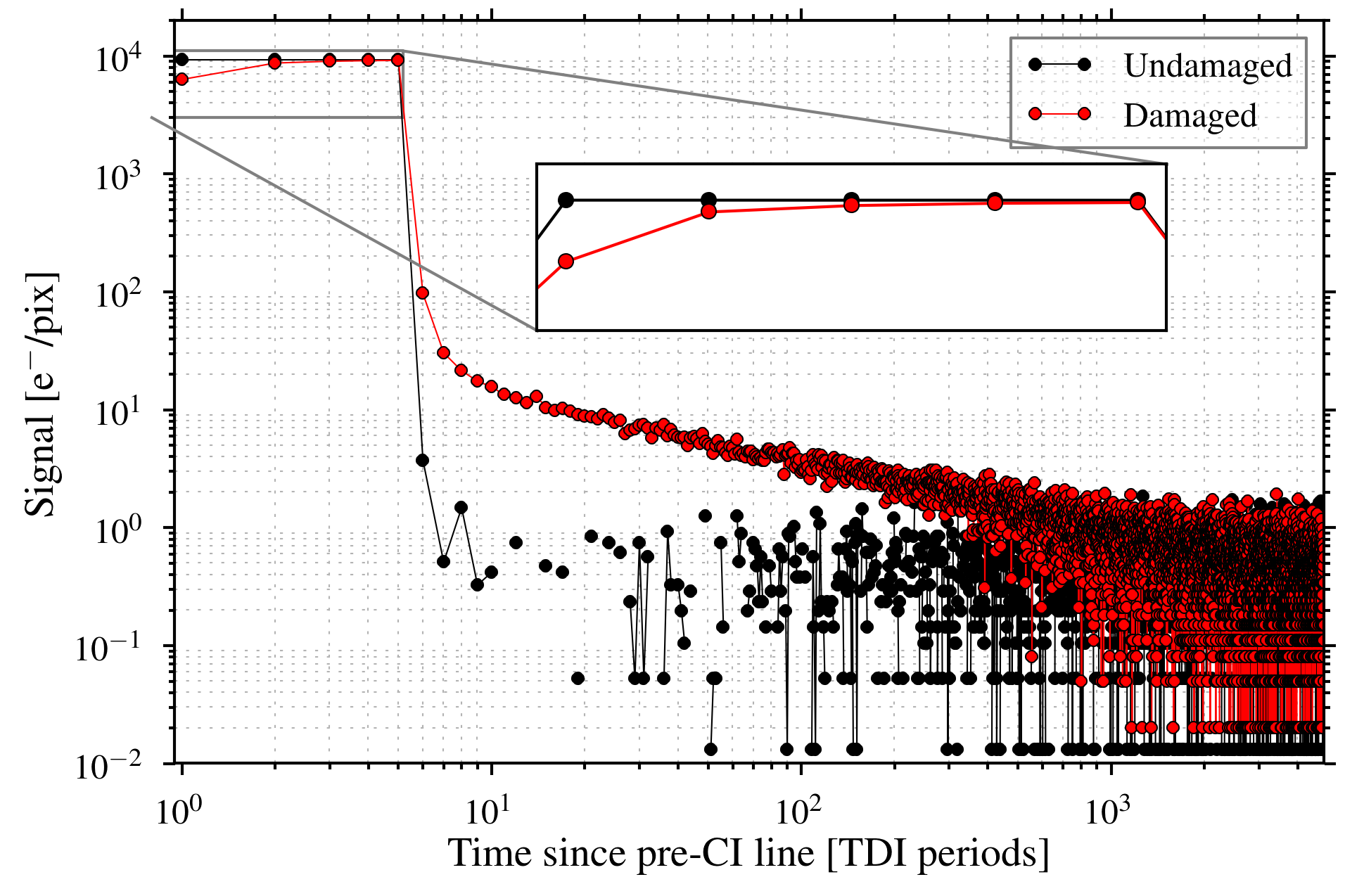}
        \caption{\label{fig:testbench}  Data acquired on the ground during the course of a pre-flight radiation test campaign on a  dedicated test-bench. Shown here are data (co-added over a number of runs) containing 5 lines of charge injection followed by $\sim5$~seconds of acquisition of the release trail. The red datapoints correspond to data acquired on a section of the detector which has been irradiated  to a NIEL of $4 \times 10^{9} ~\unit{ p^{+} / cm^2}$ of $10~\unit{MeV}$ equivalent protons. The black curve shows data acquired on an  unirradiated portion of the device for control purposes. The trapping of charge from the injected lines (the FPR) is not noticeable by eye for the black curve, but is very apparent for the data acquired from the irradiated region (see zoom in  inset).  Similarly, there is a small amount of release on the line after injection for the black curve, but a very clear and extended release trail in the red curve. }
        \end{center}
\end{figure}

\end{itemize}

\begin{figure*}
        \begin{center}
        \includegraphics[width=\textwidth]{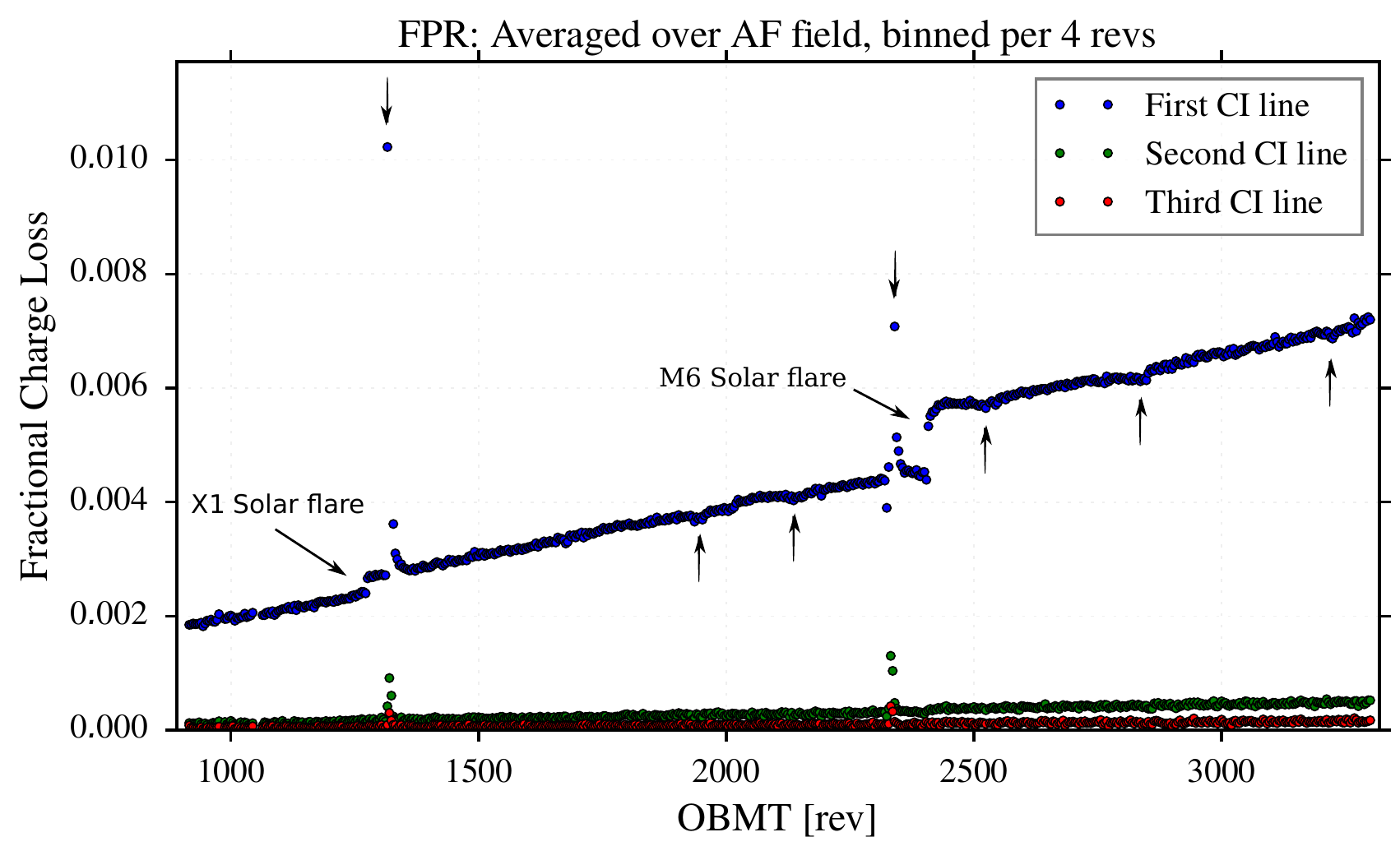}
        \caption{\label{fig:avg_n_fpr_af2-9} Averaged fractional charge loss curves for all AF2-9 CCDs, binned over 24 hours. The downward-pointing arrows indicate the times of controlled heating events where the trapping increases because  the emission time constants of slow-release traps are reduced, allowing them to empty within the charge injection period and therefore able to trap from the next block of injected lines. The upward-pointing arrows correspond to time of scans along the Galactic plane.  During these prolonged periods of high-density star regions there are more photoelectrons than normal  flooding the FPA and keeping traps filled in between charge injections, thus leading to temporary reductions in the number of electrons trapped from the  charge injection lines. Two solar proton events which clearly impacted on the FPR are also marked. The plot covers a time span of approximately 19 months.}
        \end{center}
\end{figure*}

The charge injection has a non-uniform profile in the AC direction. Indeed, the level can vary by a factor of up to~$3$  from one portion of a device to another. Thus, in Figure~\ref{fig:avg_n_fpr_af2-9} we present the number of electrons lost from the injected lines (the FPR) normalised by the injected level for that CCD column (this normalised parameter is commonly called the fractional charge loss). The FPR is calculated by assuming that the final line of injected charge approximates the true injected level in the absence of traps, and the data in the figure show the average FPR  across the astrometric field CCDs (AF2-9 CCDs) averaged over a $24$-hour time  period (four satellite revolutions). It is possible to see the effects of detector temperature increases (rises in FPR) as well as the effects of increased photo-electron generation from high stellar density sky regions (see the caption for further detail). 

Interestingly, there is a rather  linear increase in the trapping and it is possible to see clear evidence for only two step increases in addition to this. The  linear response appears to be dominated by the impact of GCRs, whilst the step increases are clearly linked to the impact of  solar proton events, as noted in the  plot (also see Section~\ref{sec:ppe_detection} for the measurement of these events in the onboard PPE rejection computations). 

Also worth noting is the  observed increase in trapping from both the second and third lines of charge. Indeed, a small amount of charge must also be trapped  from the fourth (final) line of injected charge, meaning that this method is slightly underestimating the total amount of trapping. Currently the underestimation is very small, but at some point in the future the applicability of the assumption that the final injected line is unaffected by trapping will no longer be valid. We note that for traps with release time constants much shorter  than the pixel dwell time of each charge packet, then the charge will typically be released back into the same electron packet, so these traps will not be `seen' by the FPR technique. Of course, this also means that they will not play a role in induced CTI either.

The  analysis of the de-trapping charge trails is a little more complex  than the FPR technique (owing to sensitivities on the local background/bias subtraction  and  the additional Poisson noise added to the trail from stray light, etc.), but it is found that the diagnostic produces a  similar time-dependence  to that observed in the FPR plot. 

For the science devices which have no regular periodic charge injection the CTI is monitored through onboard runs of special calibration activities that typically take place three or four times a year. For the SM devices, the nominal operation of the entire row of CCDs is temporarily suspended, the TDI gates lowered, and blocks of charge are periodically injected into the devices for some minutes. For RVS devices, the FPR and release trails are acquired as a by-product of the activity to acquire data for calibration of the CTI in the serial register (see Section~\ref{sec:ac_cti}). We do not elaborate further on the  results here, apart from noting that the trends are similar to those found for the AF, BP, and RP devices.

\subsubsection{Distribution of damage across the FPA}
\label{sec:step_increases}

\begin{figure}
        \begin{center}
                    \includegraphics[width=\columnwidth]{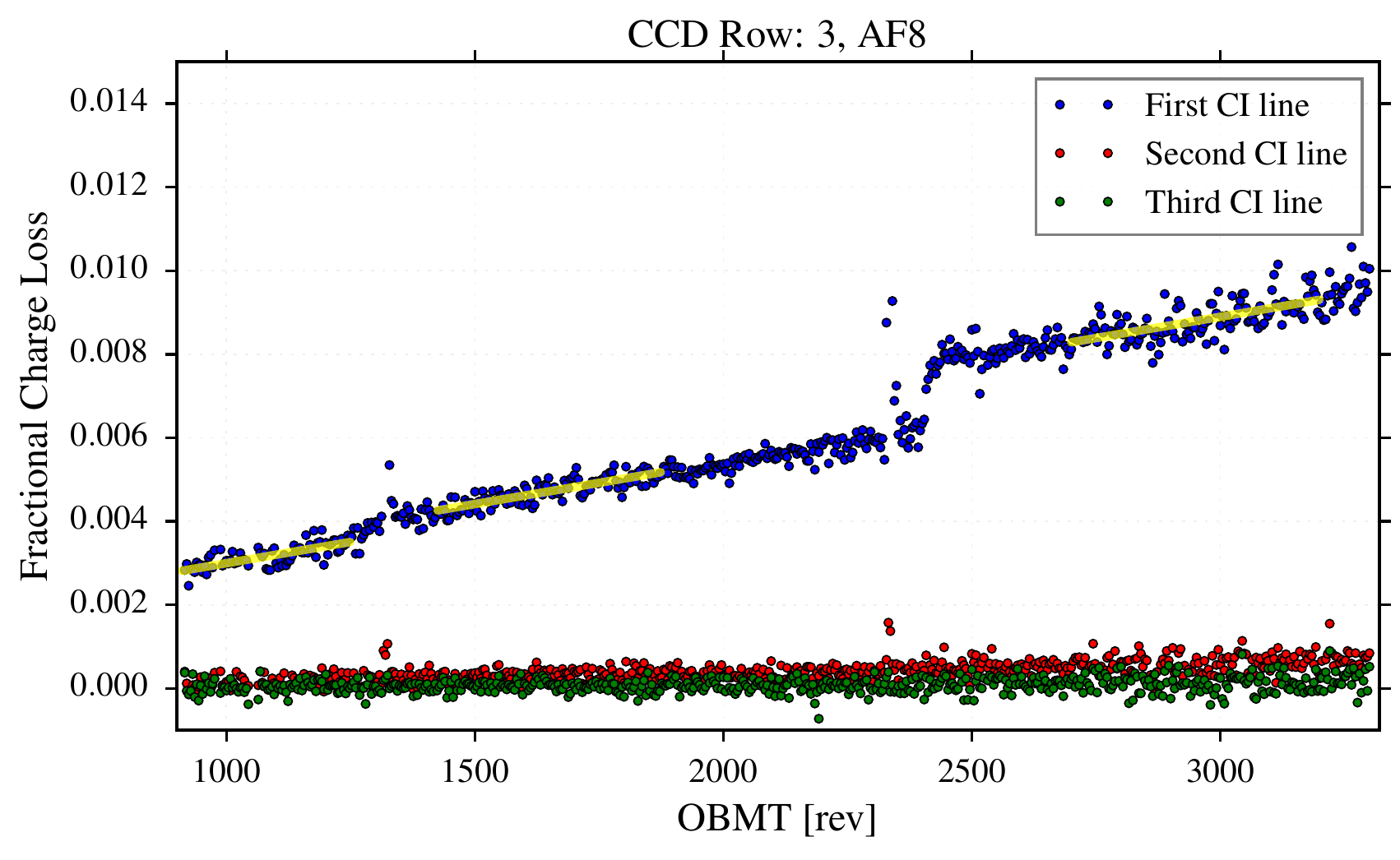}
        \includegraphics[width=\columnwidth]{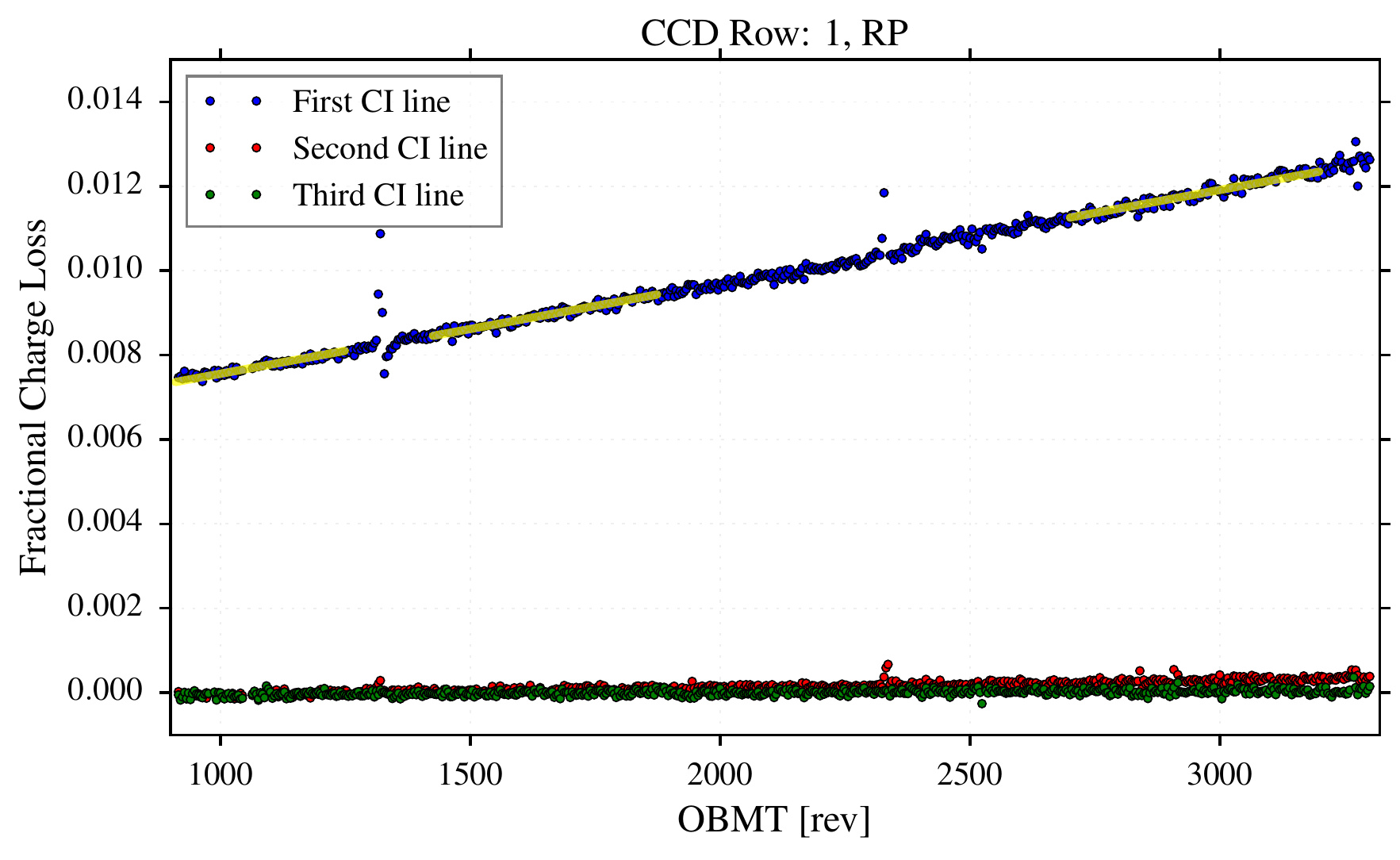}
        \caption{\label{fig:fprCcds}  {\bf{Top: }} Fractional charge loss curve for the AF8 device in row 3.  The step increases are due to the two solar proton events. {\bf{Bottom: }} Similar to top, but for the RP CCD in CCD row 1. The effect of the solar events is so low for this CCD that it is barely visible (see text for further details). Also, it can be observed that the initial FPR is higher for this CCD than for the AF CCD shown on top. This is a trait common to all of the red CCDs relative to the AF/BP devices where the measured on-ground CTI is significantly higher. Interestingly, the opposite is the case for the serial register CTI where the red devices display more efficient transfer (see Section~\ref{sec:ac_cti}). }
        \end{center}
\end{figure}

If the FPR diagnostic is evaluated on a CCD-by-CCD basis, then the geometric distribution of the variation in the level of accumulated radiation across the FPA can be analysed.  Figure~\ref{fig:fprCcds}, for example, shows  the FPR trends for two  CCDs, one close to the centre of the focal plane (the AF8 CCD in row  3) and the other towards the edge (the RP CCD in row 1). The close-to linear increase is evident for both devices (as it is for all devices), but the magnitude of the two step increases due to the solar flares is very different for both CCDs. In order to parameterise this for all devices with periodic charge injections activated, we perform a linear  fit for three different time periods. One time interval before the first flare, one after the second flare, and one in between the two (trying to avoid heating events and Galactic plane scans). The fits for each time period are overplotted in the figure.

\begin{figure}
        \begin{center}
        \includegraphics[width=\columnwidth]{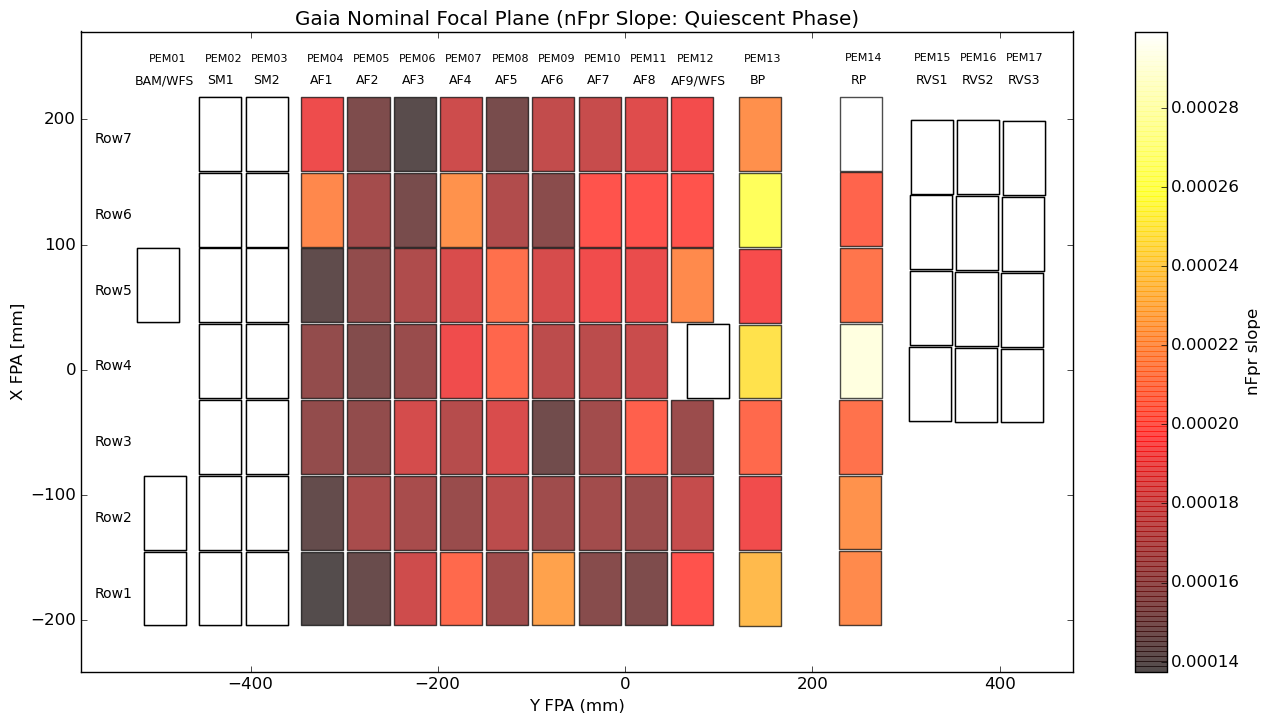}
        \caption{\label{fig:slopes}  Distribution over the focal plane of the computed  slopes of the charge loss curves between the two flare events. 
}
        \end{center}
\end{figure}

Figure~\ref{fig:slopes} depicts the distribution across the FPA of  the slopes of the fractional charge loss curves between the two solar proton events. An  examination of the slopes of each fit shows no clear pattern of variation from CCD-to-CCD, although there are suggestions that the charge loss  slopes for the BP and RP CCDs may be slightly higher than for the AF CCDs. This could be related to the longer injection period for these devices and/or secondary particles generated by the interaction of radiation with the nearby prisms; however, confirmation of this effect  requires a more detailed investigation.

\begin{figure}
        \begin{center}
        \includegraphics[width=\columnwidth]{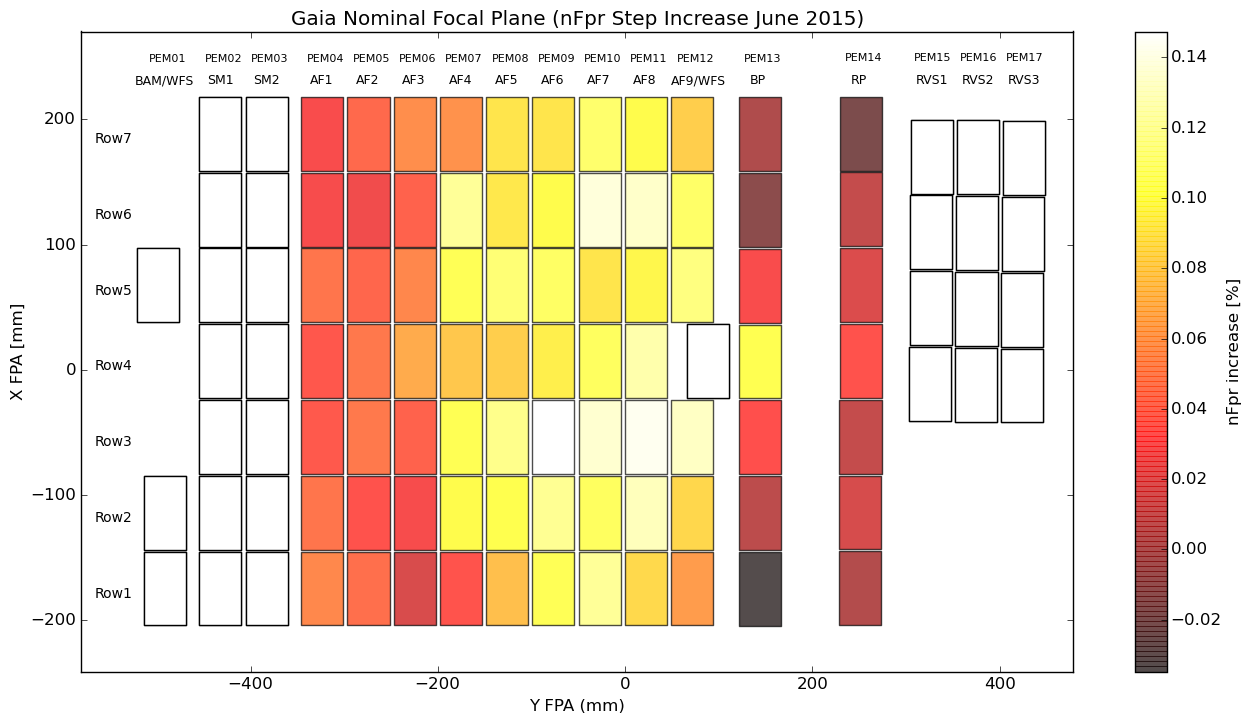}
        \caption{\label{fig:flare2}  Distribution over the focal plane of the computed step increase in the FPR due to the M6 flare of 21 June 2015. The above FPA distribution of the step increase matches the pattern predicted by pre-flight sector analysis quite well, indicating that it is likely due to the differential shielding for each device. We note that these results are only presented for AF, BP, and RP CCDs where periodic charge injection takes place continuously.
}
        \end{center}
\end{figure}

In contrast to the distribution of charge loss slopes computed between solar proton events, an examination of the geometric distribution of the FPR {{increases}} due to  solar flare events shows a clear pattern.  Shown in Figure~\ref{fig:flare2} is the response of the charge loss diagnostic to the M6 flare of 21 June 2015 (expressed in terms of percentage of injected charge that is trapped). This  flare had a   larger effect on the FPR than the X1 event of September 2014 owing  to a combination of the higher particle flux and  having proportionally more  lower energy protons than the other event (thus having a greater likelihood of interaction with the silicon atoms, rather than just passing straight through the CCD without energy loss). The damage pattern across the FPA is clearly visible and is similar to the  noisier pattern observed when analysing the earlier event. The increased effect for CCDs in the centre of the array and for   higher-order AF CCD strips is apparent, as is a rather sharp drop when moving from the strip of AF9 CCDs to the BP strip. This is due to the shielding effect of the photometer prisms. Indeed, a comparison of these results with the  end of mission NIEL predictions computed for each CCD by an industry  sector analysis carried out pre-launch  shows a strong correlation with the observed pattern. This suggests that the observed pattern is very likely due to the differential shielding effects.

\subsubsection{Comparison to pre-flight expectations}
\label{sect:cti_comparison}

In the industrial radiation sector analysis that was carried out pre-flight, an end of mission NIEL dose over the AF field of $3.11 \times 10^{9} ~\unit{ p^{+} / cm^2}$ of $10\unit{MeV}$ equivalent protons was predicted. Based on this figure, most of the on-ground radiation campaign data were acquired on devices irradiated to a fluence of  $\sim 4 \times 10^{9} ~\unit{ p^{+} / cm^2}$ ($\sim 1.2$~times higher). Through a  comparison of the FPR results presented here to those obtained on ground using irradiated devices,  we can estimate the equivalent in-flight fluence. We find that, after extrapolating the in-flight results to nominal end of mission ($\sim 6$~years after launch) and removing  the margin factor of $1.2$,  we expect  to have approximately a factor of ten less CTI at end of mission than was predicted pre-flight. We therefore note that even in the event of a mission extension,  CTI will not be the limiting factor for the mission lifetime.

 One important factor in the discrepancy between the pre-flight predictions and the observations is likely the behaviour of the Sun since launch. Both in terms of overall activity (Sun spot numbers have been low relative to the  2007 prediction) and also in terms of the number of proton events ejected towards the Earth. As has been discussed, two strong events can be easily observed through the FPR diagnostic, although  two other events  occurred very early in the mission which are apparent in the PPE rejection statistics, but which took place before the periodic charge injections were switched on. Of probably greater significance is the inherent inclusion of a margin in the sector analysis due to the  use of the industry standard $90$\% confidence levels in the study.    This margin, combined with the Sun's behaviour and the filling of traps by the very high  stray light levels present for Gaia (Gaia collaboration 2016b, this volume), can probably explain the discrepancy between the current expectation and the pre-flight predictions. We note that, although the high background levels will keep some traps filled and thus mitigate the CTI effects, it will of course add a penalty of increased photon noise which will negatively affect the astrometry, particularly for faint stars. 
Furthermore, it should be noted that the on-ground tests were carried out on devices which were irradiated at room temperature, contrasting with the much lower  temperature in flight. It is known that the irradiation temperature can affect the characteristics of the  traps and thus alter the level (and characteristics)  of the CTI effects, thus further complicating the comparison. In  fact, a study \citep{2010ITNS...57.2035H}  that compared  the effects of trapping between warm (room temperature) and cold ($133$~K) proton irradiated sections of a Gaia CCD found that the trap densities were generally around a factor of two higher for the cold-irradiated section than for the warm-irradiated section. A detailed analysis of the L2 radiation environment seen by Gaia can be found in \citet{2016arXiv160801476C}.

In any case,  the low levels of displacement damage are extremely good news regarding the quality of the   data in the Gaia data releases since it appears that the CTI-induced biases will be less than initially predicted, and the effects easier to calibrate on ground. In terms of evidence for CTI effects in the science images, some relatively crude offline analyses shows no evidence for measurable systematics of the image profile shapes when examined as a function of time-delay from the prior charge injection. For the data processing carried out for the first data release, no updates to the PSF/LSF model have been incorporated into the operation data processing pipeline (Fabricius et al., this volume) so there are currently no operational results from the pipeline on these statistics for large datasets. The CTI effects in the science images should become measurable in later data processing iterations; however, the very fact that no effects can be easily observed is good news (so far) for the science goals of the mission.
 \subsection{CTI in the serial register (AC direction)}
\label{sec:ac_cti}

The CTI during the serial register transfer, even before radiation damage, was already known  to be significant pre-flight. This is due  to the  fast serial register transfer speed  of the fast-flushed pixels ($10$~MHz) in the presence of traps that are generated during the fabrication process. At a slower transfer speed these traps become much less active, but -- owing to the TDI operation onboard Gaia -- the serial readout must be fast.  Since the precision requirements of Gaia measurements in the AC direction are less stringent than for the scan direction, this problem is not as great as it may initially appear. However, it does need to be monitored and  calibrated. In addition to the  required calibration for the distortion to the image shapes of 2D windows, the loss of deferred charge from 1D windows also needs to be fed into the photometric calibration. Therefore it is important to periodically obtain data that can be used to track and calibrate the radiation-induced degradation in the serial CTI.

\begin{figure}
        \begin{center}
        
                \includegraphics[width=\columnwidth]{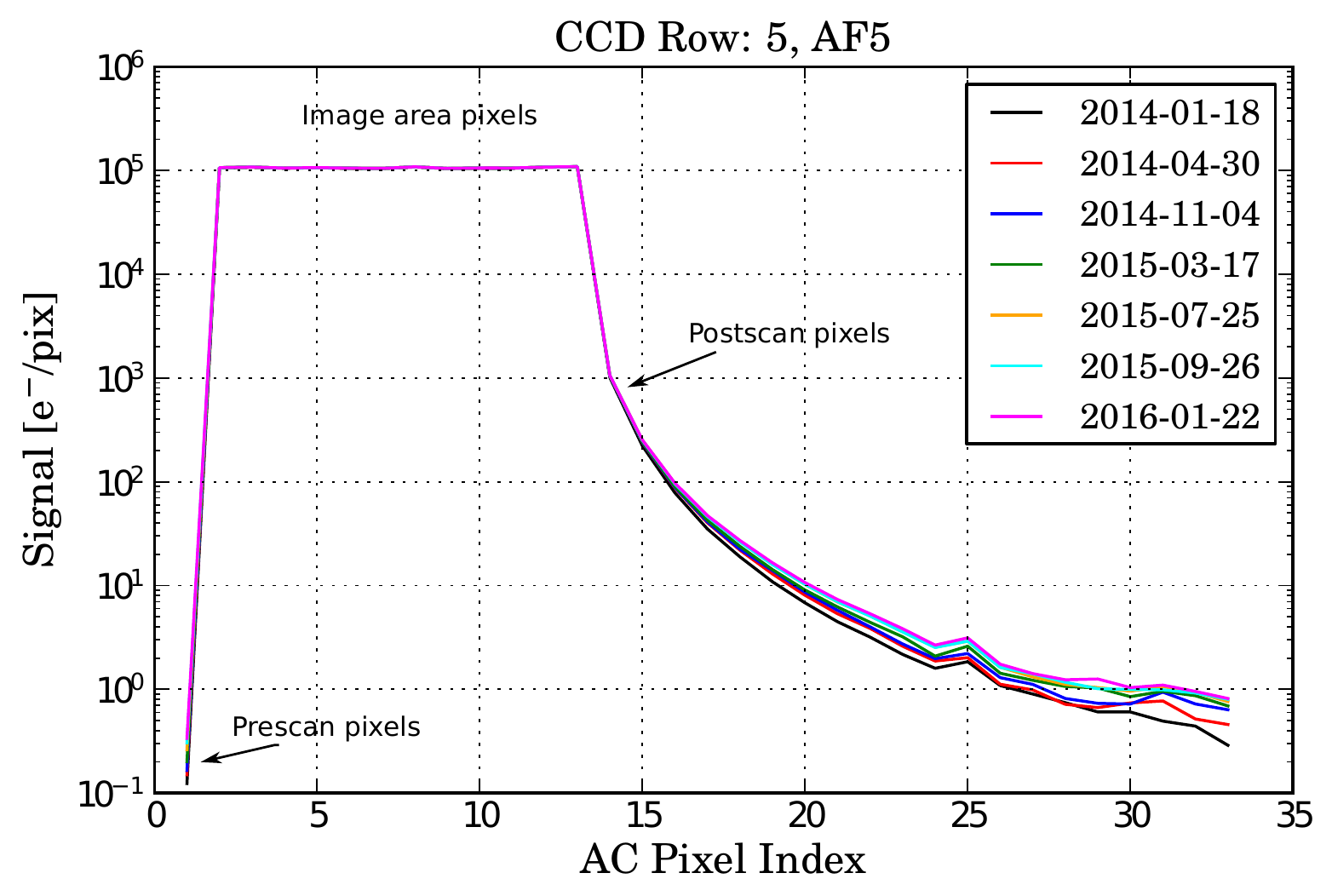}
        \caption{\label{fig:sctilevels}   Co-added and background-subtracted  AC profiles of the highest level of injected charge during each of the seven runs of the serial CTI calibration activity for a device near the centre of the FPA. The first two pixels are prescan acquisitions, followed by twelve pixels covering the final image area pixels (to monitor the injected charge level) and, finally, twenty postscan pixels to examine the trailing due  to CTI in the serial register. The slow (but measurable) degradation in the CTI is visible; also visible is the domination of the effect by native traps  present since before flight.  The date of data acquisition is contained in the legend.}
        \end{center}
\end{figure}

A standard technique for monitoring the CTI in the serial registers of CCDs is to  generate charge in the image area and then monitor any trailing into the postscan samples after  transfer through the serial register. The Gaia onboard computers are not capable of generating nominal packets containing postscan samples, so a special calibration is required to be run periodically to monitor this effect. During this activity five different levels of charge (in blocks of 255 continuously injected lines) are injected into each science device (apart from SM where the definition of the CCD operating mode precludes the acquisition of postscan samples) and VOs are placed over the end of the image area (to monitor the injected charge level) and also used to acquire postscan samples. The period of the charge injection blocks is $455$~TDI periods. To circumvent the problem where postscan samples cannot be stored in nominal packets, a special engineering functionality is used to gather the raw PEM data into special telemetry packets. It was originally planned to run this special calibration activity monthly or bi-monthly,  but since radiation damage has been lower than expected it is now run with a cadence of three to four months. After two years of  mission operations, the activity has been run seven times (including a dry run at the beginning of the commissioning phase). Shown in Figure~\ref{fig:sctilevels} are the co-added, background-subtracted data acquired for the highest level of injected charge for one particular device, and for all seven runs. It can be observed from the extent of the trails that the initial CTI was already high, but that the relative in-flight degradation  has been low (see the figure caption for further detail).

\begin{figure}
        \begin{center}
        \includegraphics[width=\columnwidth]{{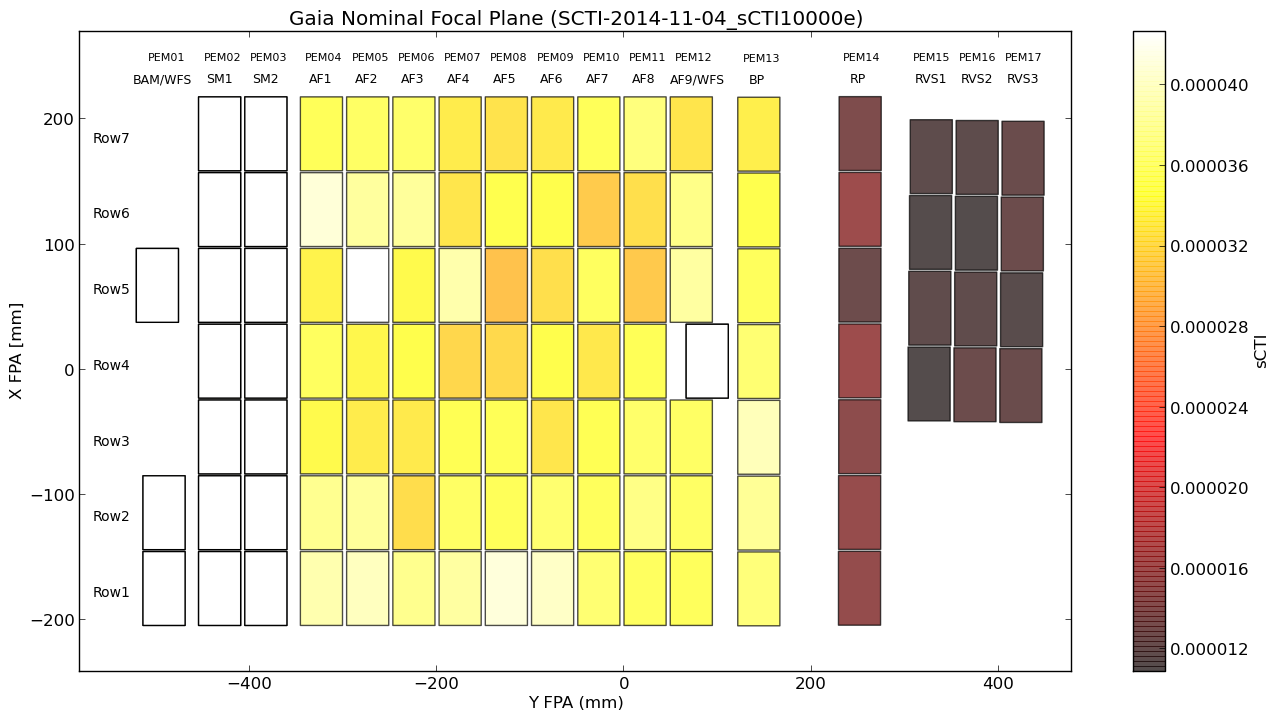}}
        \caption{\label{fig:initialscti} Close to the initial  distribution of serial CTI values (at $10~000$~electrons) as measured during the first official run of the calibration activity on board. We note the much lower inherent serial CTI for the red devices.  }
        \end{center}
\end{figure}

Shown in Figure~\ref{fig:initialscti} are  serial CTI values derived from data acquired during the first official run of the activity in  April 2014 (the first run was a dry run at a slightly higher temperature), the much lower initial serial CTI values for the red devices is readily apparent. The physical origin of the lower inherent  serial CTI for the red devices is not definitively known. However,  since the serial CTI trapping is likely due to defects consisting of silicon-oxygen complexes, it is likely that the effect is due to less oxygen contamination being able to diffuse from the optical surfaces to the channel potentials  for the thicker devices. The  values measured shortly after launch correlate well with the on-ground measurements, but -- for ease of comparison -- we show here the on-orbit values measured when the FPA temperature distribution is stable and the results can be easily compared to other runs.

\begin{figure}[ht]
\begin{center}
\includegraphics[width=\columnwidth]{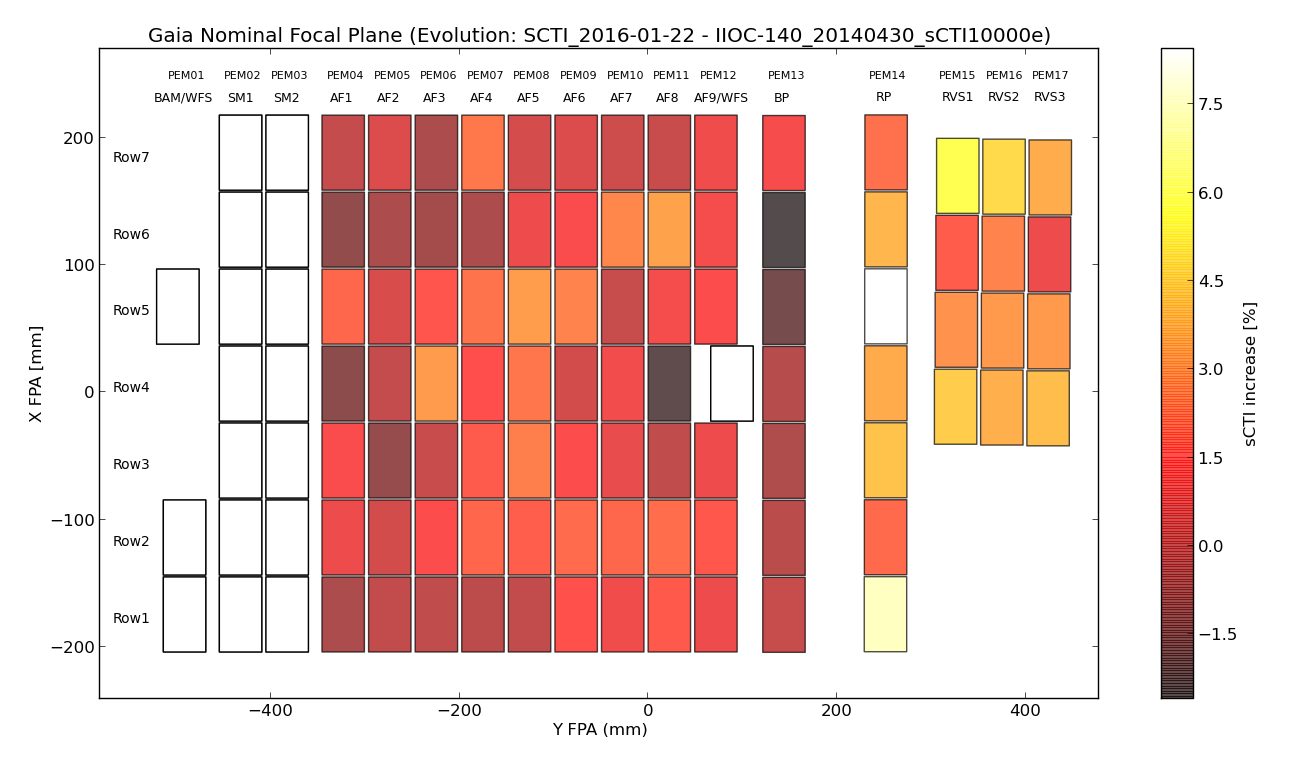}

\caption{ Distribution of the increase in serial CTI (at $10~000$~electrons) between the first official run in April 2014 and the last available run at time of writing (January 2016) in  percentage terms.
}
\label{fig:sctiincrease}
\end{center}
\end{figure}

Examination of the results for all CCDs shows that the evolution of the serial CTI is indeed steady, but is small relative to the inherent serial CTI. In order to examine the actual increase in serial CTI between the first official run of the activity  and the most recent run  (January 2016), in Figure~\ref{fig:sctiincrease} we plot a distribution of the increase in serial CTI between these two runs in terms of percentage of the initial CTI values (for a signal level of $10~000$~electrons). It can clearly be seen that the percentage increase is small for the AF  and BP CCDs ($\sim~1-3$\%) and a little higher for the RP and RVS (red variant) devices ($\sim~2-7$\%). It seems likely that this is primarily because the inherent serial CTI for the red devices was a lot lower to begin with, so the relative increase is larger.

Despite the pre-existing high levels of serial register CTI pre-flight, we note that the degradation in orbit has been modest. We also note that the findings are consistent with the lower-than-expected CTI evolution observed in the AL direction as outlined in the previous section.
Currently the acquired data is used to monitor the evolution of the CTI; however, the baseline for data processing is to use these data to calibrate a semi-physical CTI model \citep[e.g.][]{2013MNRAS.430.3078S} and to treat the raw images with this forward-model in later iterations of the data processing. 
 \subsection{Ionising radiation}
\label{sec:ionising_radiation}

Ionising radiation  causes electron-hole pair generation within the gate dielectric structures (oxide/nitride). The separated charge can become permanently trapped at the interface layers of the oxide causing a shift with respect to the applied gate voltage, commonly referred to as `flatband voltage shift'.

The CCD operating point is set to tolerate the maximum expected end of mission flatband voltage shift  of $0.5$~V, but the trend over time needs to be monitored in case steps need to be taken to avoid any degradation in the performances of the devices.  In principle, any effect can be compensated for by changing the operating voltages of the devices by the appropriate amount. A periodic run of  a special calibration activity on board is required to monitor the evolution. 

As described in  Section~\ref{sect:ciStructure}, charge is injected into each device by pulsing the injection drain with a voltage that brings the potential under the injection drain lower than the potential under the injection gate. Since the drain structure is directly connected to the silicon the applied potential is not  affected by flatband voltage shift.  This is in contrast to the potentials under the gate structures which are  modified. It  follows that there is some (CCD column-dependent) injection drain voltage at which charge is no longer injected into the device because the potential under the drain is not  low enough to inject charge over the gate. This is the turn-off voltage and tracking its evolution   permits the diagnosis of the change in the effective potentials  under  the gates over time.

 Accordingly, a special calibration activity has been developed which monitors the amount of injected charge as a function of the injection drain voltage.  The baseline operations plan is to run the activity with a cadence of $6 - 12$ months and so far it has been run on board three times. 

The average and standard deviation for the on-ground measured turn-off voltages for the 106 flight devices was $15.574 \pm 0.258$~V. A comparison with on-orbit data shows no clear  evidence for a measurable flatband voltage shift beyond the measurement noise for any one device so far. However, the mean shift over all devices between the on-ground measurement and the June 2015 on-board measurements is $+0.005 \pm 0.027$~V. The expectation pre-flight for the TID after six years at L2 was 2krad (Si), which corresponded to a shift of $\sim0.5$~V. Assuming that the TID scales with the measured flight NIEL estimates described in this paper,  after two years at L2 we would  expect to accumulate a TID of just $\sim$~70 rad, corresponding to a voltage shift of $\lesssim0.02$~V.  Therefore, after extrapolation, it is currently expected that flatband voltage shifts will not cause detector performance  issues before the end of mission. For future runs of this activity, the  sampling  of the injection drain voltages will be optimised for each CCD in order to reduce measurement  noise and better track the evolution of the flatband voltage shift.

 \subsection{Evolution of detector defects}
\label{sec:defects}

The effect of dark current, the generation of charge in a pixel  in the absence
of photons, manifests itself in Gaia
observations as `hot columns' due to the TDI mode of operation. We note that for gated observations,
 the effective integration distance is shortened, so it may be that a defect pixel far from the serial register produces  an effect in the non-gated data, but not in all or some of the gated data. Although this signal is typically
very small at normal operating temperature some columns can appear to be  very hot, producing a signal of many
thousands of ADU. Whilst most (but not all) hot columns that are being tracked on-orbit were already identified on the ground, the number and strength of such defects are expected to
increase with time.
Routine monitoring of the dark signal levels has been active in the daily processing since approximately 1600 revolutions; however, the products are only reliable
above a relatively high signal level ($\sim$20 ADU). Below this threshold there
are currently unmodelled bias non-uniformity contributions which are difficult
to distinguish from genuine dark signal (see Section
\ref{sec:non_uninformity}). 
Using this criteria, the percentage of hot columns in the combined SM, AF, and
XP strips is of the order of 0.02\%. 
However, some of these reported hot columns are not genuine, but are false detections caused by serial CTI
affecting an adjacent real hot column, and from blemish spillover.
The distribution in signal strength is presented
in Table \ref{tab:hotcoldist}, and this demonstrates the existence of defects at
all signal levels.

\begin{table}[!h]
\begin{center}
  \begin{tabular}{  c | c | c }
    \hline
    Dark signal &  Number of & Number of \\
    ADU & hot columns & affected columns \\
    \hline
    $<$20 & NA & NA \\
    20 - 100 & 7 & 16 \\
    100 - 1000 & 5 & 9 \\
    1000 - 10000 & 4 & 4 \\
    $>$10000 & 3 & 3 \\
    \hline
  \end{tabular}
\end{center}
  \caption{Distribution of hot column strength for SM/AF/XP devices at 3400
  revolutions. The centre column contains the number of real hot columns which exhibit
  increased dark signal, while the right hand column also includes adjacent
  columns which are indirectly affected via serial CTI or blemish spillover.}
  \label{tab:hotcoldist}
\end{table}

\begin{figure}
\centerline{\includegraphics[width=\columnwidth,clip]{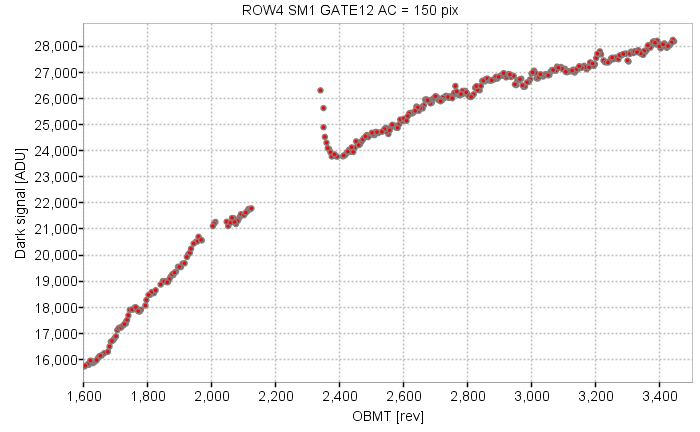}}
\centerline{\includegraphics[width=\columnwidth,clip]{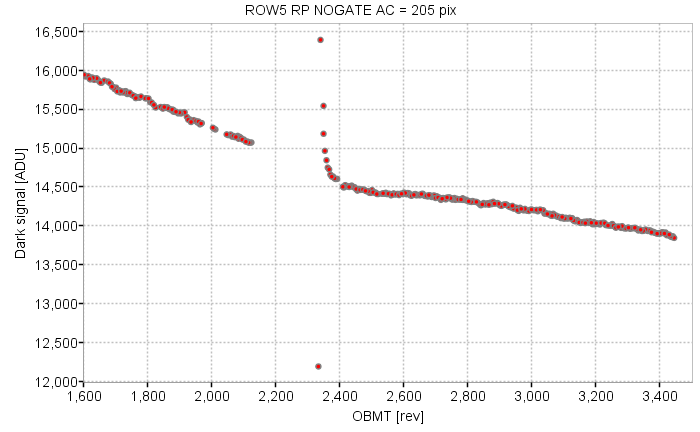}}
\caption[]{Two hot columns which
demonstrate significant variation in their strength
with time. The effect of decontamination heating
around 2300 revolutions is clearly visible.\label{fig:hotcolevolution}}
\end{figure}

\begin{figure}
\centerline{\includegraphics[width=\columnwidth,clip]{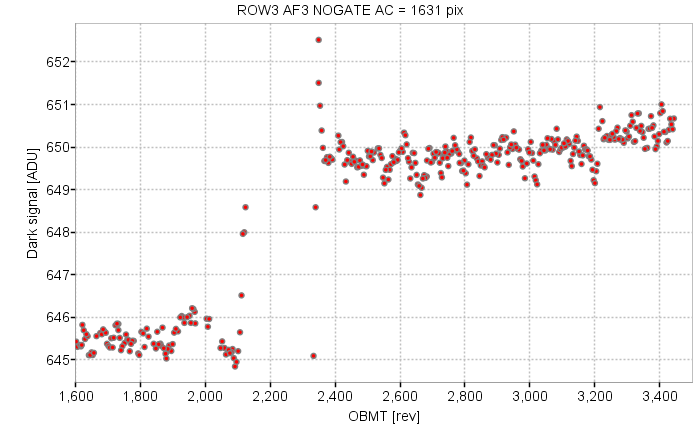}}
\caption[]{Hot column whose signal strength
has undergone a step change in strength after the
decontamination heating.\label{fig:hotcolstep}}
\end{figure}

While most of the detected hot columns show only a slight variation in their
signal level with time there are some notable exceptions. The examples shown in
Figure \ref{fig:hotcolevolution} demonstrate the significant variation in strength (increasing and decreasing) with time. In both plots the effect of the decontamination heating can been seen.
In most instances the signal from a hot column resumes its previous evolution
trend after a short interval, although we see cases where there is a step change
in level (Figure \ref{fig:hotcolstep}).

The tracking of defect columns in the  SM and AF1 CCDs is particularly
important  as this can (and does)  cause false detections/confirmations on board if not properly
accounted for in the real-time autonomous detection process. So far, only one onboard update has been required in this respect.   This was carried out in order to account for a weak ($\sim$15 ADU) radiation-induced hot column that has evolved in-flight and was causing spurious detections onboard. 

Overall, the number of radiation-induced hot columns is very low. Indeed, this  situation is expected when the short integration times for each pixel (1ms) is taken into account. For example, we can examine in more detail  the   weak column defect that was just discussed. The hot column  has evolved by $\sim$15 ADU from its initial value, which is the equivalent of $\sim$60~e$^{-}$. It follows that there must be  a pixel in that column that is generating $\sim60~000 \unit{e^{-}/\text{pix}/\text{s}}$, which would be enough to saturate the Gaia detector in just $\sim3$s if the device were operating in staring mode. The TDI operation, however, makes this hot column appear rather weak.

\section{Issues under investigation}
\label{sec:investigation}

Although all of the detectors are performing extremely well in general terms, there are  a number of relatively minor issues still actively being monitored and under investigation. 

As mentioned in Section~\ref{sec:non_uninformity}, some unexpected issues were discovered in the analysis of the bias non-uniformity activity.  These involve  on-chip cross-talk issues as well as anomalous gate release behaviour. This required some changes to the onboard implementation of the activity. In addition, other cross-talk issues, unexpected saturation behaviour, and short-term increases in the bias levels are currently being investigated in more detail.    We are also  investigating further the evolution of defect pixels, those generated in flight due to radiation processes and those  identified pre-flight.

These investigations are being performed using flight and on-ground data and are  supported through  acquisitions on  the Gaia testbench (which was used during the pre-launch radiation campaigns), which is installed and maintained at the ESA/ESTEC site in  The Netherlands. Work towards obtaining a deeper understanding of the radiation environment at L2, as well as the extraction of detector performance parameters from the results of the global calibration of the science data, are also ongoing.

\section{Conclusions}

We have presented an overview of the Gaia focal plane, the detector system, and the strategies  used for on-orbit performance monitoring of the system. We have discussed  performance results from the CCDs which are all based on an analysis of data acquired during a two-year window beginning at payload switch-on. 
We have found that the readout noise performance on-orbit of all 106 devices is excellent and that the  electronic offsets are quite stable and that, generally, any observed drifts are well correlated with focal plane temperature variations. The periodically run special calibration activity that was designed to calibrate the intra-line, readout-dependent offset non-uniformity has been successfully executed nine times so far. Despite some unexpected issues with the charge blocking TDI gates and on-chip cross talk, it is expected that the calibrations will remove the large majority of the bias non-uniformity systematics. 

We have analysed the effects of the L2 radiation environment on the devices in terms of ionising and non-ionising radiation damage, as well as the effects of cosmics on the onboard autonomous detection chain and on the science images.
Of major concern pre-flight was the  radiation-induced degradation in the CTE in the  scan direction. We have shown that the evolution of charge trapping (and release) is clearly diagnosed by the periodic charge injections which take place on most science devices. We have found that there are only two clear examples of step increases in the CTI diagnostics that can be correlated with solar proton events. Based on a comparison with FPR measurements from irradiated devices on ground it has been estimated that  CTI effects at the end of the mission will be approximately an order of magnitude less than expected pre-flight. This is thought to be due to a combination of a number of factors: (1) margins included in the pre-flight radiation sector analysis, (2) lower sun activity in general than was expected pre-flight, (3) some good fortune regarding the low number of Earth-directed proton events after the Gaia launch, and (4) the filling of electron-trapping sites by photo-electrons generated by the high stray light background. 
It is shown that the CTI in the serial register  is still dominated by the traps inherent to the manufacture process. The radiation-induced degradation in CTI, whilst clearly measurable, is of the order of $\sim~1-3$\% of the pre-flight serial register CTI for the astrometric devices and $\sim~3-7$\% for the thicker red detectors. We have also presented results from the on-orbit tracking of ionising radiation damage and hot pixel evolution and show that they should not have a significant impact on the mission science data.

Finally, we have summarised some of the on-orbit discovered detector effects that are still being investigated.

\begin{acknowledgements}

It is a pleasure to thank the Gaia Radiation Taskforce and Payload Expert groups as  well as Antonella Vallenari and Ulrich Bastian for constructive feedback and advice. In addition, we would like to thank the referee, Dr.\ Olivier Boulade, for his constructive feedback and suggestions.
We also wish to acknowledge the role of e2v technologies in the detector development process and Airbus Defence \& Space  for the CCD-PEM detector system, the FPA, and on-ground testing.

This work has made use of results from the ESA space mission Gaia, the data
from which were processed by the Gaia Data Processing and Analysis Consortium
(DPAC). Funding for the DPAC has been provided by national institutions, in
particular the institutions participating in the Gaia Multilateral Agreement.
The Gaia mission website is: \url{http://www.cosmos.esa.int/gaia}.

The authors are members of the Gaia Data Processing and Analysis Consortium
(DPAC), and this work has been supported by the following funding agencies: 

MINECO (Spanish Ministry of Economy) - FEDER through grant ESP2013-48318-C2-1-R
and ESP2014-55996-C2-1-R and MDM-2014-0369 of ICCUB (Unidad de Excelencia'Mar\'ia de Maeztu'); 
and the United Kingdom Space Agency - Gaia Post Launch Support grant.

\end{acknowledgements}

\bibliographystyle{aa} \bibliography{refs} 
\vfill
\begin{appendix}
\section{List of acronyms}
List of acronyms used in this paper. \hfill\\

\begin{tabular}{ll}\hline\hline 
\textbf{Acronym} & \textbf{Description} \\\hline
ABD&Anti-Blooming Drain \\
AC&Across-Scan (direction) \\
ADU&Analogue-to-Digital Unit \\
AF&Astrometric Field (in Astro) \\
AL&Along-Scan (direction) \\
BAM&Basic-Angle Monitoring (Device) \\
BP&Blue Photometer \\
CCD&Charge-Coupled Device \\
CTI&Charge Transfer Inefficiency \\
DPAC&Data Processing and Analysis Consortium \\
ESA&European Space Agency \\
FPA&Focal Plane Assembly (Focal Plane Array) \\
FPR&First-Pixel Response \\
GCR&Galactic Cosmic Ray\\
NIEL&Non-Ionising Energy Loss \\
OBMT&Onboard Mission Timeline \\
PEM&Proximity Electronic Module\\
PPE&Prompt Particle Event \\
PSF&Point Spread Function \\
RP&Red Photometer \\
RVS&Radial Velocity Spectrometer \\
SBC&Supplementary Buried Channel \\
SM&Sky Mapper \\
TDI&Time-Delayed Integration (CCD) \\
TID&Total Ionising Dose \\
VO&Virtual Object \\
WFS&WaveFront Sensor \\
\\\hline
\end{tabular} 
\end{appendix}

\end{document}